\documentclass{article}
\pdfoutput=1
\usepackage[utf8]{inputenc}
\usepackage{authblk}
\usepackage{setspace}
\usepackage[margin=1.25in]{geometry}
\usepackage{graphicx}
\usepackage{subcaption}
\usepackage{amsmath}
\usepackage{float}
\RequirePackage{amsfonts}
\usepackage{multirow}
\usepackage{longtable}
\PassOptionsToPackage{hyphens}{url}
\usepackage{hyperref}
\usepackage{url}
\usepackage{xcolor}
\hypersetup{
    colorlinks,
    linkcolor={blue!80!black},
    citecolor={blue!50!black},
    urlcolor={blue!80!black}
}
\begin{document}
\sloppy
\title{\textbf{Humanode} \\

  \small{Whitepaper v. 0.9.6 ``You are [not] a bot''} }
\date{\today }

\author{Dato Kavazi, Victor Smirnov, Sasha Shilina, MOZGIII, MingDong Li, Rafael Contreras, Hardik Gajera, Dmitry Lavrenov, and the Humanode Core}
\maketitle
\begin{abstract}

  The pursuit of new decentralized financial networks has engulfed the world.
  In the last few decades, dozens of different protocol prototypes have been deployed
  to achieve a decentralized state of finance, but most are unable to overcome
  plutocratic governing systems that derive from the principles upon which
  Proof-of-Work (PoW) and Proof-of-Stake (PoS) heavily rely.
  The advent of blockchain technology has led to a massive wave of different
  decentralized ledger technology (DLT) solutions.
  Such projects as Bitcoin and Ethereum have shifted the paradigm of how to
  transact value in a decentralized manner, but their various core technologies have their
  own advantages and disadvantages. This paper aims to describe an alternative to
  modern decentralized financial networks by introducing the Humanode network.
  Humanode is a network safeguarded by cryptographically secure bio-authorized nodes.
  Users will be able to deploy nodes by staking their encrypted biometric data.
  This approach can potentially lead to the creation of a public,
  permissionless financial network based on consensus between equal human nodes
  with algorithm-based emission mechanisms targeting real value growth and proportional emission.

  Humanode combines different technological stacks to achieve a decentralized,
  secure, scalable, efficient, consistent, immutable, and sustainable financial network:

  \begin{itemize}
    \item a bio-authorization module based on cryptographically secure neural
          networks for the private classification of 3D templates of users' faces

    \item a private Liveness detection mechanism for identification of real human beings

    \item a Substrate module as a blockchain layer

    \item a cost-based fee system

    \item a Vortex decentralized autonomous organization (DAO) governing system

    \item a monetary policy and algorithm, Fath, where monetary supply reacts to real
          value growth and emission is proportional

  \end{itemize}
  All of these implemented technologies have nuances that are crucial for the integrity
  of the network. In this paper we address these details, describing problems that might
  occur and their possible solutions. The Humanode core acknowledges the power of
  liveness detection and internal multimodal biometric processing methods that,
  implemented properly, will tremendously increase resistance against Sybil attacks
  and overcome the challenges and limitations of modern biometric authentication and
  identification systems. The main goal of Humanode is to create a stable and just
  financial network that relies on the existence of human life.
\end{abstract}
\newpage

\tableofcontents
\newpage

\section{Introduction}

\emph{From truth = authority to truth = money, how much has changed?
  In the Humanode protocol, truth = human existence}

Increasing attention to decentralized ledger solutions over the recent decade gave rise to a
whole class of projects oriented toward encryption methods and consensus mechanisms.
These scientific pursuits have never been exposed to capital in such a direct manner.
An abundance of capital has created a massive research wave on a variety of decentralized
transaction verification systems. Simultaneously, biometric processing has evolved to a stage
where search and matching operations can maintain privacy while liveness detection error
probabilities decline by the day.

Humanode is a network based on cryptographically secure bio-authorized nodes.
Using solutions that provide private search and matching operations and liveness detection algorithms,
users will be able to deploy nodes to create a public permissionless financial network based on
consensus between human nodes who share the fees and ownership of the network in an equal manner.

Modern decentralized verification systems rely on the concept of material obligations to prevent
malicious activity—Proof-of-work (PoW) blockchain systems blacklist mining equipment,
Proof-of-Stake (PoS) slashes tokens. The main focus of these protection mechanisms is
to create a system in which attacks are unimaginably costly for any hypothetical predator.
This reliance rises from problems of distrust on many levels, but most importantly because
any trust system requires an instrument for preventing malicious activity. As human nodes are
not created through mining farms or monetary obligations in the form of staking,
they are not exposed to the same angle of attack. The Humanode network will prevent malicious
activity by blacklisting biometric data, meaning that your biometric identity becomes the stake.

Human nodes are created through cryptobiometric authentication, which is a combination of
cryptographically secure matching and liveness detection mechanisms to verify the uniqueness and
existence of real human beings. Thus, the user's pseudonymous biometric identity becomes the stake
that gives access to the creation of a node that verifies transactions.
This approach mitigates the problem of the disproportion of power in decentralized systems such as
mining cartels or validator oligopolies. In the Humanode network, only one node derives from one
biometric identity. This also means that every node is equal in terms of voting and computation power,
while rewards for verification and storage are equally distributed among the human nodes.

As the right to launch human nodes is not entangled with a native token,
it allows the implementation of any monetary system without the necessity of conforming to the
requirements of capital-based Sybil-defense systems. With human nodes replacing staked assets,
it is now possible to avoid a disproportion of token emission between those who stake, validate,
or simply hold the asset.

Humanode will implement the Fath hypothesis as the mechanism for monetary supply adjustment
and proportional distribution of emission. The main idea behind the Fath hypothesis is a
full-reserve system that calculates the amount of goods and services sold in equal periods of time.
If the value created in the new period is greater than the value in the previous one by 1\%,
the Fath protocol issues 1\% of the supply and delivers it to every single wallet in the network,
depending on the account balance (savings). If the wallet holds 1\% of the supply during the emission,
it gets 1\% of the minted tokens directly from the protocol.

Any person in the world, no matter where they are from or who they are,
can become a human node, as long as that person has access to devices that can conduct
biometric processing (for example, a smartphone with a camera and biometric processing
applications for recognition) or other verified hardware. The system delivers the equality
of every single human node by deriving only one node from one biometric identity and mitigates
any disproportion of power due to reward equality of individuals. As the system implements
the Fath hypothesis, which negates the effect of devaluation on agents of the system,
this narrows gaps between the users of the network as the emitted value is distributed
proportionally to every participant.

The main goal of Humanode is to create a stable and just financial network that relies
on the existence of human life itself. We aim to alleviate all the intermediaries that
stand between a person and his ability to become a validator of transactions. Humanode strives
to deliver easy node creation flows and make it natural for any human to verify their unique
existence privately in any digital service. We acknowledge the power of a strong and idea-driven
community, and that is why Humanode will be an open-source project. We believe that by joining
forces together with passionate minds and hearts throughout the world we will be able to achieve
a balanced state of the system that will ensure our economic freedom and stability and safeguard
the future of our children, grandchildren, and many generations to come. The symbiosis of
humans and technology is inevitable, and Humanode is just a small but important step in the
large transcendence period that we are all going through.

\section{Humanode network}

The Humanode network is a protocol that can prove one's unique identity through private
biometric authentication schemes and grant permission to launch a node and verify
transactions running a public permissionless network based on collective human existence.

The core stack of technologies enabling human node technology include:

\subsection{Cryptobiometric blockchain protocol}

\subsubsection*{Substrate}

\begin{itemize}
  \item Single node per person invariant (bio-authentication consensus)
  \item EVM compatibility
\end{itemize}

\subsubsection*{Cryptobiometric neural networks}

\begin{itemize}
  \item Zero-Knowledge proven active and passive liveness Detection
  \item ZK-proven encrypted face feature extraction
  \item Distributed encrypted feature search and matching
\end{itemize}

\subsubsection*{Distributed economy}

\begin{itemize}
  \item Cost-based network fees
  \item Fath as the monetary policy and algorithm targeting real value growth and proportional emission.
\end{itemize}

\subsubsection*{Humanode/Vortex DAO}

\begin{itemize}
  \item Proposal pool system
  \item On-chain voting ``Vortex''
  \item Formation
\end{itemize}

\section{Humanode’s goals and objectives}

In our research, the Humanode core came to agree that the Humanode network should have two
main epochs that are defined by two factors, the external existence of keys and how often
proof-of-human-existence occurs. In our opinion, these two factors also shape the way
people accept becoming a part of the Humanode network.

The first factor is derived from the problems connected to presentation attacks.
If we consider that every year 3D (or even 4D) printing becomes more accurate,
then sooner or later models will be able to emulate floating points even for top-notch neural
networks and will not be expensive to produce.
Since the very first creation of keys as a tool to store something that you do not
want to easily fall into the hands of strangers or malicious actors, keys have always
existed externally. Even when humanity learned to digitize keys, we still used some
virtual plane of existence to store them. Modern biometrics are also based on keys
stored externally because most companies use external modalities (iris scans,
fingerprints, nose, palm, ears, etc.). Malicious actors may try to steal your biometric data,
using a futuristic printer to bypass the biometric processing protocol. We can mitigate this
angle of attack by removing the key from the external world per se and putting it inside a human body.
There are a lot of ways to do so, but the most common are biochemical, DNA signatures,
and brain-computer interfaces (BCIs). If the network is built so that node creation is
possible only through internal biometric processing, where keys do not exist externally,
then the mechanism becomes tremendously impenetrable. At some point, Humanode will have to
transcend to this kind of verification of human existence. Before the implementation of internal
biometric processing protocols is complete, Humanode will use external multimodal biometric
processing with liveness detection.

The second factor consists of two main sub-factors: real proof-of-human-existence and
fee distribution. Let us imagine that somehow we create tech that emulates internal
biometric processing with such accuracy that it bypasses the system and is able to
generate an artificial proof-of-human-existence. If the Humanode network demands verification
every month then a malicious actor would be able to create many identities without additional
challenges in terms of bandwidth and computing power. But if the Humanode network requires
proof of existence every, let us say, half a minute, then it would be very hard and costly
to carry out such types of attacks, considering that the biometric processing protocol itself
has an incorporated neural network that distinguishes emulations from real data by detecting
liveliness. The second sub-factor is all about fees. Fees are built in such a way that they
are equally distributed among every node that exists in the network. If Humanode requires proof
every month, then if some human, unfortunately, ceases to exist, his node would still receive
rewards until the end of that month.

Humanode’s long-terms goals and objectives:

\begin{itemize}
  \item Sybil resistance based on decentralized pseudonymous biometric identities

  \item secure, scalable, efficient, consistent, immutable, and sustainable Substrate mechanism

  \item creation, proliferation, and development of a strong and dedicated community of human nodes

  \item custom low-latency high-throughput Sybil-resistant consensus protocol

  \item privacy-preserving biometric processing protocols

  \item distributed encrypted biometric templates matching

  \item ZK-proven liveness check

  \item EVM compatibility with the Substrate pallet

  \item native Humanode applications (wallets, etc.)

  \item biometric ownership;

  \item integration with EVM-compatible applications

  \item rich integration into Substrate ecosystem

  \item Humanode token (HMND) with Fath monetary system, equal fee distribution, and proportional emission distribution on the Humanode network

  \item Vortex, a DAO that regulates the existence of the Humanode network through voting

  \item a proposal system that pulls trending proposals to vote in Vortex

  \item Formation, a system that distributes grants across approved proposers and helps them to assemble a team to develop Humanode solutions

  \item a public Humanode knowledge base that stores all the information, research and analytics carried out by teams assembled in Formation. It will also act as a base in educational sessions and programs carried out in Humanode

  \item cryptobiometric applications that deliver proper flows in UX/UI

  \item multimodal biometrics

  \item ability to deploy Fath-based monetary systems

  \item ability to deploy your own tokens on Humanode with different monetary policies and systems

  \item Humanode development framework with modular solutions

  \item judicial system framework;

  \item BCIs for liveness detection

  \item neurosignatures for signing on-chain transactions

  \item internal modalities-based biometrics (neurosignatures, DNA matching)

  \item real-time proof-of-human-existence

  \item biometric-based Autonomous Intelligent Agent

  \item ZK offchain commitment-based smart contracts
\end{itemize}

\subsection{Solutions for governments}

The reliance of human beings on governments is undisputed.
We believe that many countries are willing to try and make their financial
networks more secure, decentralized, and just. So we are willing to help out
any administration or any other form of government that is willing to deploy
human node–based national currencies, backed by citizen human nodes or even a
Fath monetary system for fair, direct, and proportional distribution of emission
and extraction of excessive monetary supply. Fees generated through verification
of a user’s unique existence can also become a solution in terms of funding a
universal basic income. Governments that decide to make this experiment and pursue
a transcendent solution will get the full support of the international human node
community, and the Humanode core will assist those courageous people with research,
analytics, and development, if Vortex, the Humanode network decision-making body, approves.

\subsection{Shout-out to white hats}

As we are building a system that is based on highly experimental data,
we want to ask all the white hats to try and pwn our network hard.
The future security of a network is very dependent on the amount of pwnage
it has to go through in the early days of its testing and creation. Someone
finds an angle of attack, tests it, shares the results with the Humanode community,
and makes a proposal to research and develop a solution that mitigates this attack.
Then, if this proposal is approved by Vortex the team behind the solution gets a
grant with rewards for pwnage. So please come and test our system’s resolve,
we will get stronger with each attack.

\section{Cryptobiometric blockchain protocol}

Humanode implementation is planned to be a maintainable, reliable codebase
that will be under active development in the years to come. To build a high-quality system,
a good foundation is needed, and the first thing to settle on is a
programming language for the implementation.

\subsection{Rust}

We have evaluated Rust, Go, and C++ and chose Rust for our implementation,
with Go being the second-best option. Rust offers a very expressive type system and a novel
approach to memory safety, ultimately being the best choice for a long-term project with
high demands on code quality and maintainability.
Being a language that compiles to native code rather than a VM, it is also very efficient.
Rust follows C++'s principle of “if you don't use it, don't pay for it” and powers
developers with zero-cost abstractions. It doesn't use a garbage collector,
and even async runtime is implemented in the user code, rather than being a part of the language.
By modern standards, this is a very high degree of control.
The downsides of Rust are that it is difficult to master, and the development process
with Rust is slower than with, for instance, Go. Rust enables you to take control
over many aspects of a system, and it takes knowledge and time to implement them right.
We figured that for our use case, the benefits far outweigh the cost, and thus, we chose Rust.

\subsection{Substrate}

Every blockchain developer faces a choice: build the codebase from scratch or adopt an existing one
for their needs. This decision is simple on the surface, but it is a tough one to resolve.

Substrate is a modular framework that enables the creation of purpose-built blockchains
by composing custom or pre-built components. Since dismantling the involvement of the
token in the consensus mechanism is one of our main goals as a project,
we want to focus our development on building biometric-based consensus aspects on the network.
And that is why the modular customization vision is a great ally on our path.

We chose Substrate after carefully evaluating the alternatives
(build the code from scratch and using one of the other existing codebases) for several reasons:

\begin{itemize}
  \item Substrate is designed to be used as a platform to build blockchains, i.e.,
        it is by design a developer tool rather than a final product.
        Using it was more appealing than basing our development on a codebase that
        implemented a particular blockchain project.

  \item Substrate code quality is rather high, and it is clear that people working on it
        care about the quality.

  \item We decided that we want something that is a library rather than a framework,
        and with Substrate, the flexibility and modularity of the code are exactly
        where we needed them. If we were building the code from scratch,
        we'd go with a very similar approach to the one that Substrate developers have taken.

  \item Substrate is being constantly worked on and improved by many people and this
        means that we can get a stable stream of improvements simply by building on Substrate.
        There is also a vibrant community around Substrate and many people we can talk to
        that can help us if we face any trouble with Substrate itself.

  \item  Substrate has proven to be an excellent tool so far, and it allows us to
        focus on issues specific to Humanode, rather than spending time on the common
        ones of blockchain building.

\end{itemize}

\subsection{Components}

Here’s an overview of our system on the component level.
Only a subset of the most important pieces are represented here, and there’s a lot more
that wasn’t included.

Below is the list of key components that covers the main Humanode requirements to achieve our goals.

\textbf{Humanode App}—an application that allows users to be a part of the Humanode network
by exploring the network, submitting transactions, passing biometric identification,
running a node, joining the block production, participating in the Humanode DAO.

\textbf{Humanode Peer (substrate-based node)}—a blockchain node in the Humanode network.

\textbf{Humanode-runtime}—a business logic that defines Humanode network behavior including storage,
state transition logic, block, and transaction processing.
Also, it enables one of the defining features of Substrate-based blockchains: forkless runtime upgrade.

\textbf{Humanode-rpc}—a component that allows blockchain users including the Humanode app
and other dapps to interact with the Humanode network by HTTP and WebSocket RPC servers.

\textbf{Consensus}—a logic that allows Humanode network participants to agree on the
state of the blockchain defining one of the key Humanode features: proof-of-human-existence.

\textbf{Aura consensus}—a deterministic consensus protocol that primarily provides
block authoring by a limited list of authorities (validators) that take turns creating blocks.
The authorities must be chosen before block production begins and all authorities must
know the entire authority set. Time is divided up into “slots” of fixed length.
During each slot, one block is produced and the authorities take turns producing blocks in order forever.

\textbf{Grandpa consensus}—deterministic consensus protocol that provides block
finalization where each authority participates in two rounds of block voting.
Once two-thirds of the authorities have voted for a particular block, it’s considered finalized.

\textbf{Bioauth consensus}—an addition to deterministic consensus protocols that is
responsible for validating whether block authors of proposed blocks have successfully
passed biometric authentication (bioauth-authorized).

\textbf{Frontier consensus}—a component that provides an Ethereum compatibility layer
that allows the running of Ethereum dapps natively by enabling the functionality of
running EVM contracts, Ethereum block emulation, and transactions validation.
\newline\newline
\textbf{Main Pallets (runtime modules)}
\newline

\textbf{Bioauth}—a component that defines required storage items and calls to manage
authentication tickets that allow bioauth-authorized peers to participate in block production.

\textbf{Fath}—responsible for providing a Humanode monetary algorithm with a proportional
distribution of issued tokens.

\textbf{Cost-based fee}—provides a logic to enable Humanode transaction fee economics of the network.

\textbf{Ethereum}—a module, combined with the RPC module, that enables Ethereum block emulation,
validates Ethereum-encoded transactions, and allows existing dapps to be deployed on a
Humanode network with minimal modifications.

\textbf{Biometric Provider}—a component that provides biometric registration and authentication
that guarantees the image’s privacy and resistance to various vectors of attack on biometrics.

The scheme below illustrates communication and interaction between the components.

\begin{figure}[H]
  \centering
  \label{fig1}
  \includegraphics[width=0.75\linewidth, keepaspectratio]{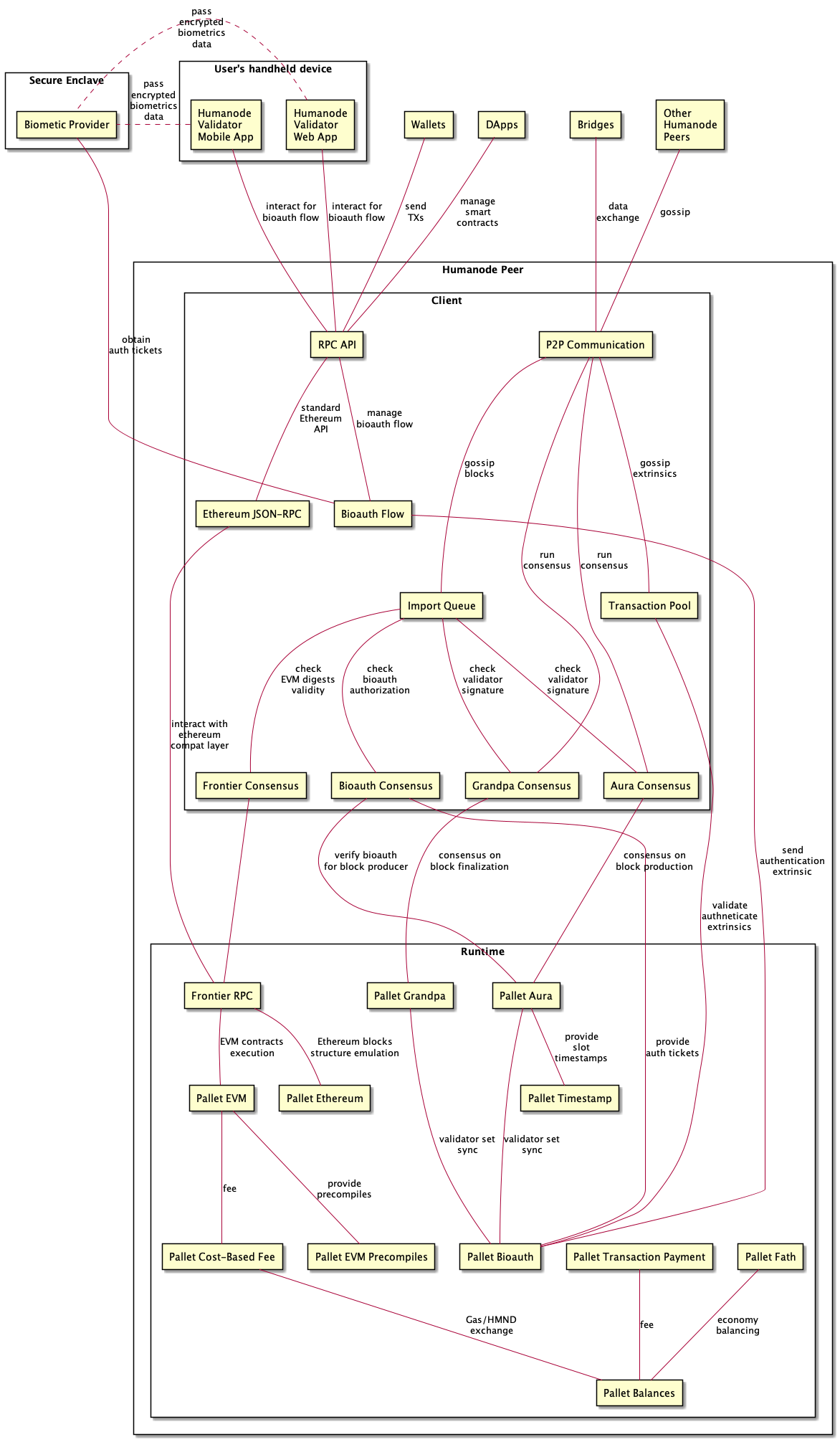}
  \caption{Humanode network components}
\end{figure}

\subsection{Consensus agnosticism}

One of the interesting features of Humanode that we pursue is consensus agnosticism,
the ability to change the consensus mechanism of the network if the Humanode DAO approves it.
It derives from the necessity for constant research on the most suitable consensus
for a leaderless system with equal validation power of nodes. Different consensus
mechanisms have various pros and cons which constantly evolve, change, and shift
due to the large amount of research done by thousands of scientists across the world on this topic.
Swappable consensus mechanisms would allow the Humanode network to evolve and not
be constrained by a framework of a singular consensus. Moreover,
the Substrate ecosystem has ongoing attempts (\cite{Swappable}, \cite{Re-Genesis}) to support such a feature.

\subsection{Residual increase in the number of human nodes}

The limitation on the number of human nodes will be set following Vortex voting.
When the limit is still low, it remains possible to attack the system by coordination
among active human nodes. In order to lower the number of malicious human nodes that
verify transactions in the early days of the protocol, we impose additional selection
criteria for candidate human nodes based on Proof-of-Time and Proof-of-Devotion.
Those with a higher tier or longer governing history are going to be the first candidates
to enter the Humanode public network on the main net as nodes.

\subsection{Ethereum compatibility}

To bring cryptobiometric technology to existing protocols, the Humanode network
includes an EVM pallet that allows it to run Solidity smart contracts and use
existing developer tools. Its implementation is based on SputnikVM, which consists
of four modules: evm, evm-core, evm-runtime, and evm-gasometer.

One of the goals of the Humanode network is to solve current issues of transaction
fee pricing by making the fees stable in USD terms despite the volatility of the
native token. Hence, the evm-gasometer is replaced with a cost-based fee system.

Unified accounts, first proposed by Moonbeam, solve the problem of account
incompatibility between H256 Substrate addresses and H160 Ethereum addresses
where the user is unable to send transactions directly and thus have to have two
accounts and move assets between them to get access to both chains. With unified accounts,
a single H160 is all the user needs to make the multichain experience seamless.

Moving a dapp or a smart-contract framework from Ethereum to Humanode will require minimal changes.
Solidity smart contracts, block explorers, development tools, bridges, and frameworks
for decentralized autonomous organizations are easily ported to the chain based on equal validators.

By bridging Humanode to other EVM-compatible chains, the network will be able to
provide private biometric processing and Sybil resistance to dapps and protocols
based on other chains. A biometric smart contract written
in Solidity deployed on a needed chain will communicate with, for instance,
a decentralized finance protocol, and then send the request to the Humanode network
where biometric data is stored. Without revealing the user's identity, the Humanode
network sends ZK liveness proof and identity checks back to the DeFi protocol to prove
the user is the same real human being without using any PII (Personally Identifiable Information).

\subsection{Slashing system}

Any verification system requires executive tools that safeguard the consistency,
immutability, and governing mechanisms of the network. The fiat credit cycle utilizes
law enforcement and jails, PoW blacklists mining equipment, PoS slashes staked cryptoassets,
and Humanode slashes your biometric identity by blacklisting malicious
actors for a period of time. If a malicious actor tries to harm the network
in any way, the system will blacklist his biometrics. After being slashed,
the perpetrator will remain blacklisted for a determined period and will not
be able to sign any transactions in the Humanode network. That period is
defined by the severity of the malicious act itself. Any changes to the
severity levels of perpetrations and blacklisting periods are defined through Vortex.
Proposal rights to change any slashing conditions are given to Governors upon reaching the Legate tier.

Some perpetrations have blacklist-period scaling mechanisms.
Blacklist periods start with the base parameters stated in the table below
and then can only scale upwards. Scaling has steps predetermined by
Vortex: 0.5 months (the basic length), 1 month, 2 months, 3 months, 6 months,
1 year, 2 years, 3 years, 10 years, 20 years, and forever.

\begin{table}[H]
  \centering
  \caption{Slashing severity levels and types.}
  \label{table1}
  \resizebox{\textwidth}{!}{
    \begin{tabular}{|p{5cm}|c|c|p{5cm}|c|}
      \hline
      \textbf{Perpetration}                                                & \textbf{\vtop{\hbox{\strut Severity}\hbox{\strut level}}} & \textbf{\vtop{\hbox{\strut Blacklisting}\hbox{\strut period (months)}}}
                                                                           & \textbf{Additional consequence}                           & \textbf{\vtop{\hbox{\strut Blacklist-period}\hbox{\strut scaling}}}                  \\
      \hline
      Did not verify existence once in a month                             & 0                                                         & 0.5                                                                     &
      Human node is excluded from validator set and stops receiving fees from the protocol
                                                                           & No                                                                                                                                               \\
      \hline
      Made a proposal where the meaning of propositions does not match the proposal type
                                                                           & 1                                                         & 1                                                                       & None & No  \\
      \hline
      Failed to deliver upon an agreed Formation proposal in time          & 2                                                         & 1                                                                       & None & Yes \\
      \hline
      A node remained offline for more than 48 hours                       & 2                                                         & 0.5                                                                     &
      Human node is deactivated and stops receiving fees from the protocol & Yes                                                                                                                                              \\
      \hline
      Made a proposal where the meaning of propositions did not match the proposal
      type and the Governor himself did not have the right to propose that type
      because of the tier level                                            & 3                                                         & 1                                                                       &
      Human node is deactivated and stops receiving fees from the protocol & Yes                                                                                                                                              \\
      \hline
      Node overall uptime less than 91\%                                   & 3                                                         & 1                                                                       & None & Yes \\
      \hline
      Tried to push in a false transaction                                 & 5                                                         & 120                                                                     &
      Human node is deactivated and stops receiving fees.
      Governing time and proof-of-dedication nullified                     & Yes                                                                                                                                              \\
      \hline
    \end{tabular}}
\end{table}

\section{Fath}

Humanode dismantles the involvement of tokens in the consensus mechanism,
meaning that different monetary systems can be implemented on top of the
Humanode network without the necessity to conform with the requirements of
token-entangled protocols. Humanode will implement the Fath hypothesis as
the basis for the circulation of HMND (the Humanode Token).

Fath is a monetary algorithm with a proportional distribution of issued tokens.
The amount of issuance is determined by the amount of additional value created in
the monetary system—the economic output of goods and services sold. The distribution
of issued tokens happens proportionally based on the currency savings of each holder.
When the output of the economic system around Fath currency rises by 1\%,
1\% of the monetary base is issued. As a result, every wallet gets 1\%
of the currency on top of its holdings.

The idea behind Fath is to create a monetary system where emission is distributed
proportionally, in contrast to how modern fiat credit-cycle financial networks and
capital-based public blockchain networks operate.

\begin{figure}[H]
  \centering
  \label{fig2}
  \includegraphics[width=0.75\linewidth, keepaspectratio]{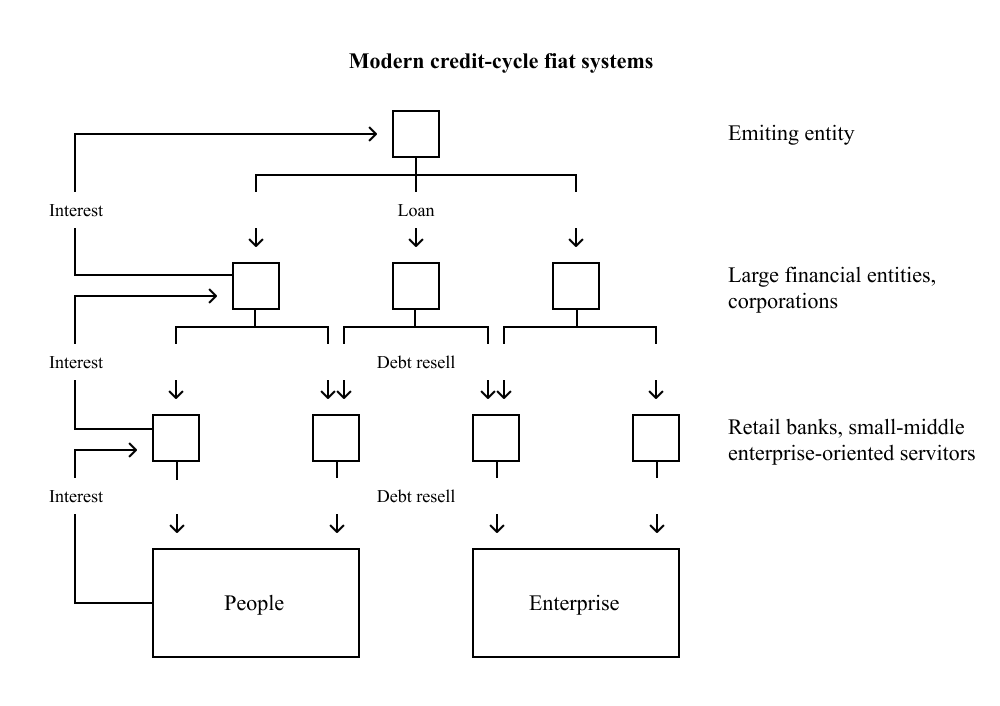}
  \caption{Issuance in modern credit-cycle fiat systems.}
\end{figure}

With the global conversion to fiat and decimalization that overwhelmed most
countries in the early 1970s, world leaders decided to transcend us all to a system
in which emission is injected as a form of debt. Afterward, it is passed down the
system in the form of loans. Even if we leave out the fact that some of that issuance
forever resides on one of the upper levels because of corruption and fraud, people,
enterprises, and retail banks are the ones who are constantly cornered because they
are the ones paying for those emission and the only ones they can resell their debt
to is each other. If for some reason one of the large financial organizations fails
to accumulate enough money to cover its expenses and interest then in most cases the
emitting entity prints a relief package to save it. If ordinary people or enterprises
fail likewise, in most cases they are fined, thrown onto the street by law enforcement,
go bankrupt, or go to jail. Consider the fact that every time the emitting entity prints
money it increases the monetary supply and devalues the currency, meaning that agents
at the bottom of the emission pyramid not only get devalued with each coin printed,
they also pay for it to happen.

\begin{figure}[H]
  \centering
  \label{fig3}
  \includegraphics[width=0.75\linewidth, keepaspectratio]{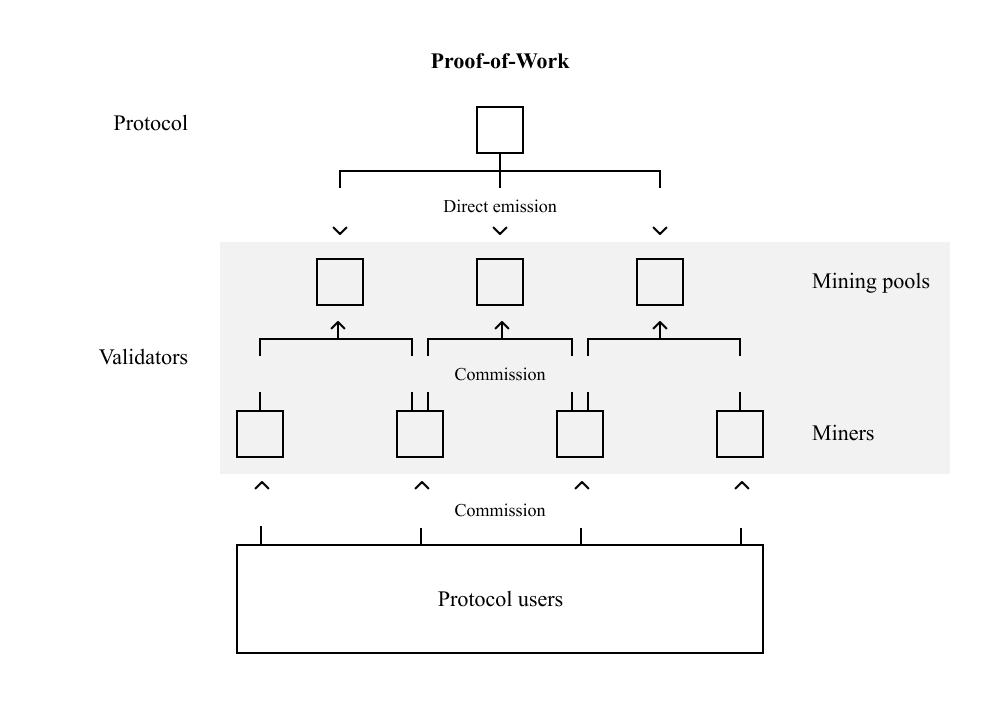}
  \caption{Issuance and commission in PoW blockchains}
\end{figure}

In PoW blockchains, the protocol acts as the emitting entity. Most PoW coins have set
the emission and max supply. For example, Bitcoin has a max supply of 21 million coins.
At the time of the creation of this paper, its circulating supply is 18.8 million.
With emission set in every block and the halving that happens every four years, it
will take approximately 120 years to mint everything. Emission is received by miners
not in the form of a loan, but directly. However, only miners receive it.
Ordinary users and even financial entities that hold large chunks of Bitcoin get nothing.
Miners either decide to hold onto the emitted money or sell it on the market.
This system does not sell debt to the agents at its bottom, but devaluation of non-miner
agents’ assets, even if ridiculously small, still happens, as the emission is
received only by miners. Another thing is that supply is not balanced with value creation,
meaning that the limited supply does not line up with the growth of value in the system.
That makes it deflationary, which on a nation-sized scale makes economies unhealthy and
can even lead to a crisis.

\begin{figure}[H]
  \centering
  \label{fig4}
  \includegraphics[width=0.75\linewidth, keepaspectratio]{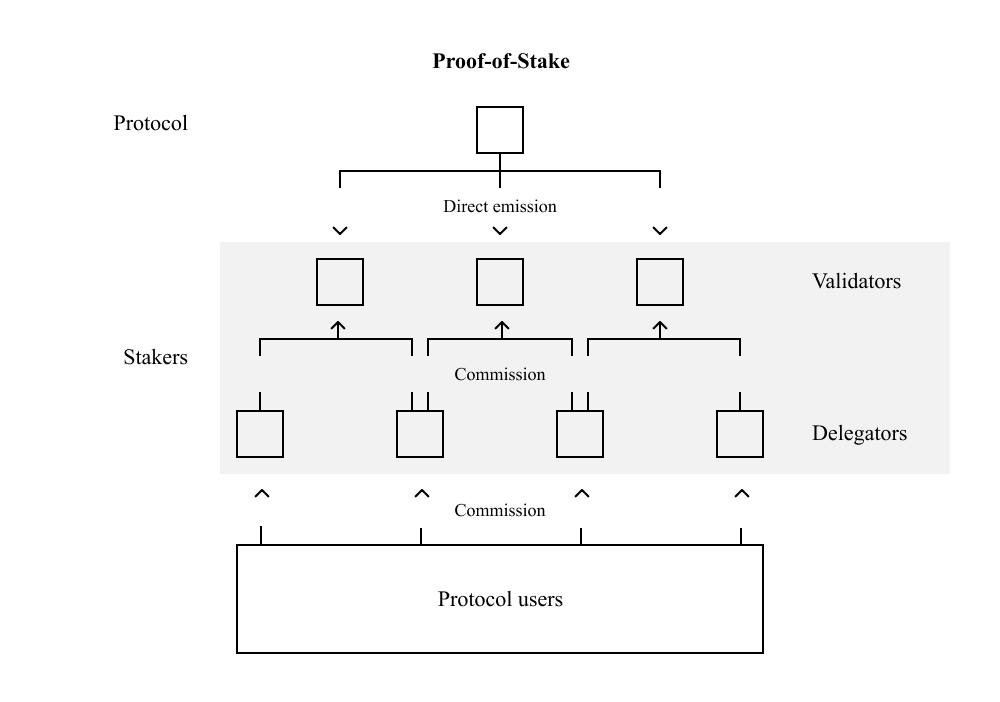}
  \caption{Issuance and commission in PoS blockchains. }
\end{figure}

As in PoW, in PoS the protocol acts as the issuance entity.
In most cases, PoS have some kind of a governing entity that decides upon emission;
it can be either pre-set as in Bitcoin or it can be flexible with many different
methods of realization. Commonly there is a DAO that sets the emission. As in PoW,
validators receive issuance directly from the protocol, but in delegated PoS,
they also redistribute it across their Delegators. Protocol users get nothing
from emission and DAO can set emission at any level. Sometimes devaluation is
very strong because validators accumulate minted tokens and sell them on the market
to cover expenses and for profit—at the same time, their networks are not as big as
Bitcoin, which counterweighs the devaluation effect.

\begin{figure}[H]
  \centering
  \label{fig5}
  \includegraphics[width=0.75\linewidth, keepaspectratio]{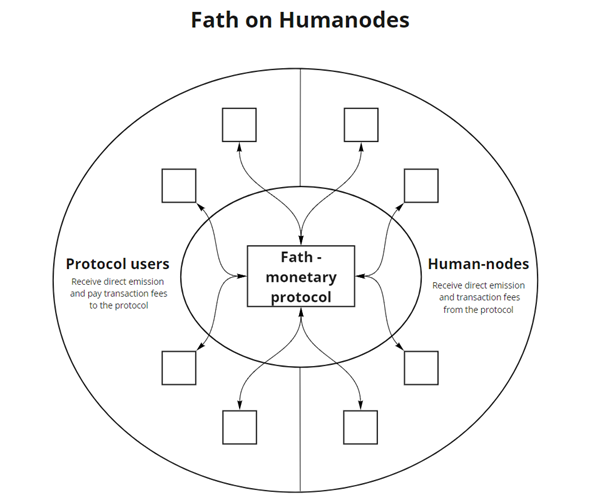}
  \caption{Issuance and commission in Fath}
\end{figure}

The emission of tokens in Fath behaves differently from the systems mentioned above.
One of the hypotheses that are the basis of Fath is that it is possible to mitigate
the long-term effects of devaluation by the proportional distribution of emission.
Emission is delivered to every single member of the network directly from the protocol,
regardless of whether a person is a validator or not.

The amount of emission is defined by the Fath protocol algorithm,
which calculates the difference between real value creation (Gross Network Product; GNetP)
in two different time periods. If GNetP in the second period is different from GNetP
in the first then the algorithm calculates the difference and changes the monetary
supply by the same percentage.

We consider the HMND token first of all to be a transaction-processing as well as a
biometric network, which is why GNetP in the first implementation of Fath will be
calculated based on the fees spent by participants of the network. If the amount of
commission received by human nodes in the second period is different from the first,
then the algorithm applies the same difference in percentage to supply and rebalances
every single wallet that exists.

Two types of rebalances occur, inFath and outFath:
\begin{itemize}
  \item If the amount of commission paid out in the second period of time
        exceeds the commission paid out in the first period, then inFath occurs
        and emission is distributed across every wallet proportionally
  \item If the amount of commission paid out in the second period is smaller
        than in the first, then outFath occurs and the protocol proportionally burns
        excessive supply throughout every single wallet as well
\end{itemize}

\subsection{Transaction-based emission algorithm example}

\textbf{End of Year 0 }

Supply: 10,000,000 HMND

Commission paid out: 1,000,000 HMND

Your wallet: 1000 HMND

\textbf{End of Year 1 }

Supply: 10,000,000 HMND

Commission paid out: 2,000,000 HMND

As commission paid out in Year 1 exceeds the same quantity in Year 0 by 100\%,
inFath occurs. A total of 100\% of the supply is minted and given out to
everyone proportionally to ledger balances.

New supply: 20,000,000 HMND

Your wallet: 2000 HMND

\textbf{End of Year 2 }

Supply: 20,000,000 HMND

Commission paid out: 1,500,000 HMND

As commission paid out in Year 2 is smaller than in Year 1 by 25\%,
outFath occurs. A total of 25\% of the monetary supply is burned and rebased proportionally.

New supply: 15,000,000 HMND

Your wallet: 1500 HMND

Such a rebalancing mechanism tries to:
\begin{itemize}
  \item Mitigate the long-term effect of devaluation due to disproportional emission
  \item Negate macroeconomic shocks and structural inefficiencies that occur due to the
        monetary supply not satisfying the needs of the growing or shrinking GNetP.
        If you are interested in Fath hypotheses that are based on data from monetary
        systems from the 3rd century BC, you can read more about them here.
\end{itemize}

\subsection{Implications of the transaction-based Fath system for a distributed network economy}

If the total fees in the Humanode network in the terms of the dominant
currency stay the same, the price change of the HMND token will be
followed by changing the price of the transaction in HMND according to the
formula stated in \emph{Fee-setting and distribution mechanism in the Humanode network.}

As we found out, the fee paid in HMND changes opposite to the price change.
The token issuance is tied to the change of total fees collected by the protocol.
If the changes in quantity and size of the transactions are bigger than
the asset price change, only then the protocol will initiate inFath.

However, over time, the network and its token obtain new use-cases other
than trust in processing valuable data. That is when we need to account
for value creation in the network and derive a value that was created in the system,
other than transaction processing. When the system obtains new properties,
the new Fath modules should be launched to account for new values created a
nd to change the algorithm accordingly. In the end, Fath is supposed to have
modules that combined are capable of self-accounting for as many transaction
and contract types as are involved in the calculation of GDP.

\section{Biometric approach to user identification }

Rapid development in IT, DLT (Distributed Ledger Technology), and AI are prompting
biometrics to constantly innovate and make the most of market demand. According to
the latest reports, the global biometrics market is forecast to reach \$82.8 billion \cite{BiometricsGlobalMarket-88.8} to
nearly \$100 billion \cite{BiometricsGlobalMarket-100} by 2027, growing at a $>19.3\%$ compound annual growth rate (CAGR)
from an estimated \$ 24.1 billion in 2020.
According to these reports, the multimodal biometric systems segment is projected
to increase in revenue at a significant CAGR during the forecast period.

In terms of authentication type, voice recognition is supposed to witness
significant growth due to consumer desires for a safer identity mechanism.
Facial recognition is also poised for growth, as it is witnessing a boost
from the launch of Apple’s Face ID system \cite{Apple-faceID}.

In 2020, the global market for mobile biometrics was estimated at \$18 billion,
and it is projected to reach a  size of \$79.8 billion by 2027, growing at a CAGR of
23.7\% over the analysis period 2020–2027 \cite{MobileBiometricsGlobalMarket}. Growth in the scanner segment was
readjusted to a 20.1\% CAGR for the next seven-year period.

Furthermore, the post-COVID 19 global digital identity verification market is
forecast to grow from \$7.6 billion in 2020 to \$15.8 billion by 2025, at a CAGR of 15.6\%.

The ability to privately secure user authentication through biometrics has
been the goal of many cryptographic researchers. For the last two decades,
cryptographers have concentrated their efforts on solving the problem of
biometric protection against malicious activities of the verifier.
Solutions like biohashing, biometric cryptosystems, and cancelable biometrics
have not evolved enough to become efficient and secure for a hypothetical
user (\cite{Davida1998} ; \cite{Ratha2001}, \cite{Ratha2002};
\cite{Jin2004}; \cite{Kong2006}; \cite{Teoh2008};
\cite{Rathgeb2011}; \cite{Syarif2014}; \cite{Nair2015}).

Until not so long ago, biometric identification methods carried a heavy risk
to personal privacy. Biometric data are considered to be very sensitive,
as they can be uniquely associated with a human being. Passwords are not
considered PII, as they can be changed and not associated with any person directly.
The main risks associated with biometric matching in the past were based on the
fact that they required the biometric data to be visible at some point during the process.

\subsection{Humanode bio-authorization overview}

The privacy and security of biometric data have been among the most critical
aspects to consider when deciding on a biometric and cryptographic technology
to use in Humanode. Biometric registration and authentication are carried out
through a novel method based on cryptographically secure neural networks for
the private classification of images of users' faces so that we can:

\begin{itemize}
  \item guarantee the image's privacy, performing all operations without the
        biometrics of the user's face having to leave the device
  \item obtain a certificate or proof that the operations are carried out correctly,
        without malicious manipulation

  \item have resistance to different attacks, such as the Sybil attack and reply attack

  \item carry out all registration and authentication operations without the need
        for a central entity or authority that handles the issuance and registration of users' cryptographic keys

  \item compare the feature vector each time the user wants to authenticate in a cryptographically secure way
\end{itemize}

Let's now see how the different technologies that we use to perform the
registration and authentication of users are broken down, guaranteeing
privacy in a decentralized environment.

Traditionally, neural networks are used to identify an image.
A neural network is a particular case of a machine-learning technique that consists
of a series of so-called nodes structured in layers. These nodes or
neurons are mathematical functions that perform a specific operation
according to the layer they belong to.

For example, the convolutional layer is in charge of filtering the
information to determine the similarity between the original image
covered by a filter and the filter itself.
The activation layer also determines if the filter pattern defined
in the convolutional layer is present at a particular position in the image.
There is also a layer called max pooling that modifies the data to make them easier
to handle \cite{Kikuchi2008}.

When the user logs into the system for the first time, the neural network gives
us a unique feature vector that identifies the user. Once this vector is registered,
we can store it for future comparisons when the user wishes to authenticate.

The main objective of the biometric registration and authentication system is
to protect the images of users throughout the whole process and on the
different layers of the neural network.
It is required that the operations are carried out effectively and efficiently,
preventing unauthorized access to the data, from the time when they are obtained
on the user's device to when they are processed in the neural network
and registered in the system \cite{Belguechi2011}.

A malicious user gaining access to the neural network should
not be able to obtain any sensitive information.
This is why Humanode biometric system architecture is designed to
run neural networks locally on the user's device and only send the
proof that all the neural network layers were executed.
The user will also send the neural network's output in the
form of an encrypted feature vector.

\subsection{Convolutional neural network}

Often referred to as CNNs or ConvNets, convolutional neural networks
specialize in processing data that are grid-like in topology, such as images.

In a digital image, each pixel contains a binary value that denotes
how bright it is and what color it should be.
It contains a series of pixels that are arranged in a grid-like format.

\begin{figure}[H]
  \centering
  \label{fig6}
  \includegraphics[width=0.75\linewidth, keepaspectratio]{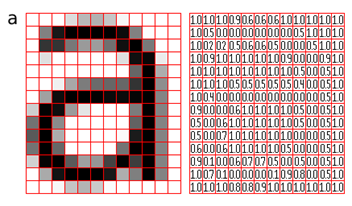}
  \caption{Representation of image as a grid of pixels (\cite{pixels}).}
\end{figure}

Each neuron works in its own receptive field, interconnected with other
neurons so that the entire visual field is covered. The human brain processes
enormous amounts of information as soon as it sees an image.

In the same way that each neuron in the biological vision system responds
to stimuli only in its receptive field, each neuron in a CNN also processes
information only within its receptive field. With a CNN, one can enable
computers to sense simpler patterns (lines, curves, etc.) at the beginning
and more complex patterns (faces, objects, etc.) as they progress.

There are four main layer types of CNNs: a convolutional layer,
pooling layer, fully connected layer, and one or more activation layers.

\begin{figure}[H]
  \centering
  \label{fig7}
  \includegraphics[width=0.75\linewidth, keepaspectratio]{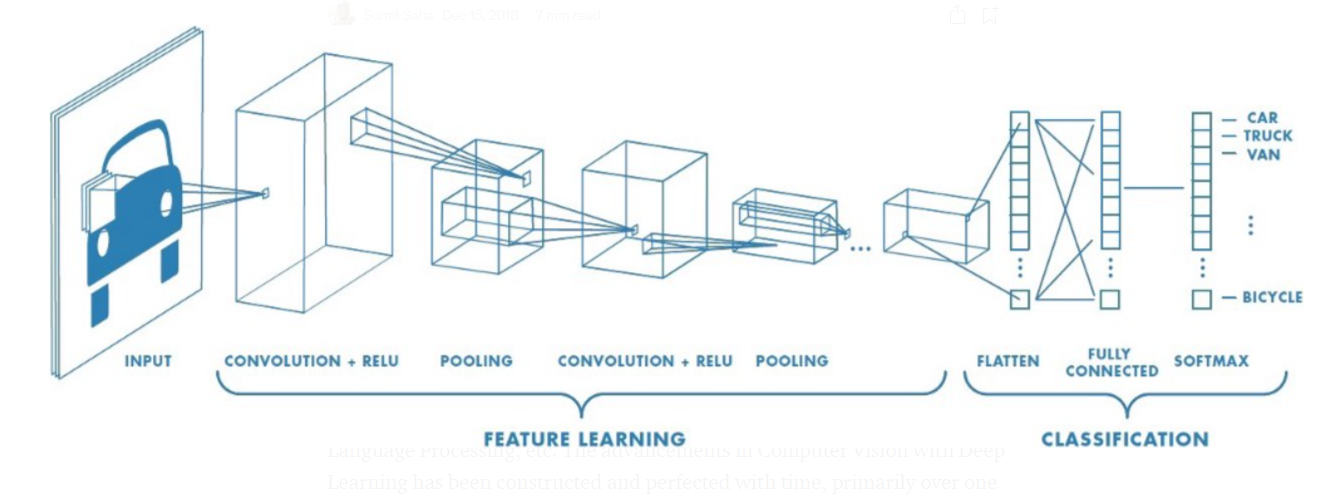}
  \caption{Architecture of a CNN (\cite{CNNarch}).}
\end{figure}

\subsubsection{Convolution layer}

CNNs have a convolution layer that carries a vast amount of computations.
Using this layer, we perform a dot product between two matrices, one
that contains the set of learnable parameters, known as a kernel,
and the other that contains the restricted portion of the receptive field.

In the case of an image composed of three (RGB) channels, the kernel
height and width will be smaller than the image, but the depth will
encompass all three channels.

When the forward pass is made, the kernel slides across the height
and width of the image, creating an image representation of the
receptive region. A kernel response is generated by computing an
activation map in two dimensions, which results in a representation
of the image for each spatial position. A stride refers to the size
of the kernel as it slides. The size of the output volume can be
calculated as follows if we have an input of size $W \times W \times D$ and a
number of kernels of size $F$ with a stride $S$ and a padding $P$:

$$W_{out} = \frac{W - F + 2P}{S} +1$$

This will yield an output volume of size
$W_{out} \times W_{out} \times D_{out}$.

\begin{figure}[H]
  \centering
  \label{fig8}
  \includegraphics[width=0.75\linewidth, keepaspectratio]{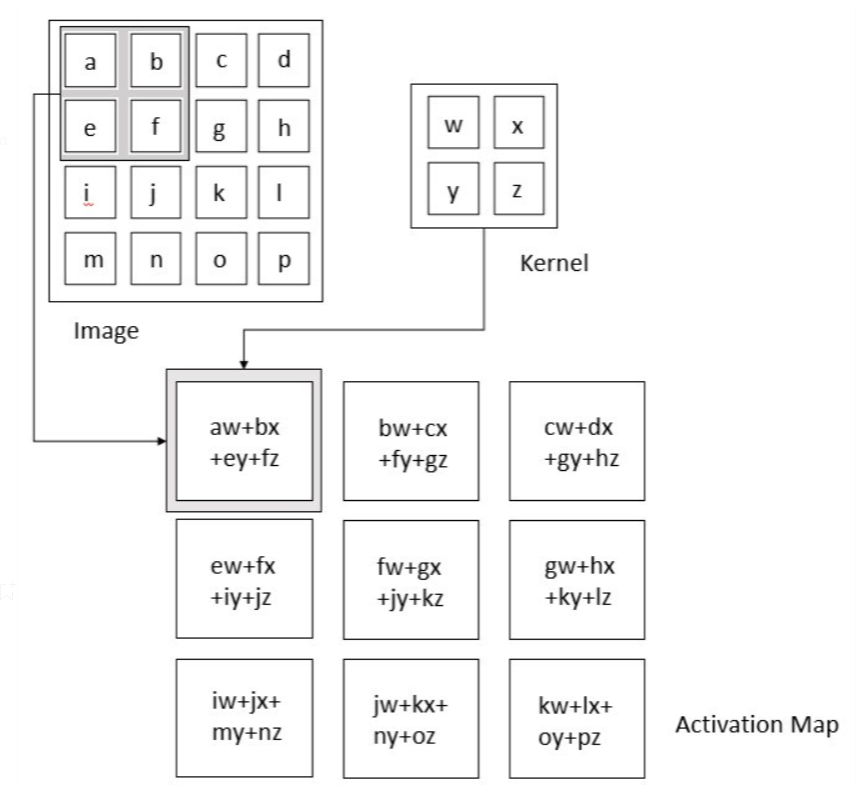}
  \caption{Convolution Operation (\cite{gholamalinezhad2020pooling}).}
\end{figure}

\subsubsection{Pooling layer}

During the pooling layer, summary statistics are derived from the nearby
outputs to replace certain outputs of the network. As a result, the size
of the representation is reduced resulting in a decrease in computation
and weights. The pooling operation is applied to every slice in turn.

In addition to the rectangular neighborhood average,
there are several pooling functions such as the $L2$ norm of the rectangular
neighborhood and the weighted average based on the distance to the
central pixel. Max pooling, however, is the process most commonly
used, which reports the max output from the neighbors.

\begin{figure}[H]
  \centering
  \label{fig9}
  \includegraphics[width=0.75\linewidth, keepaspectratio]{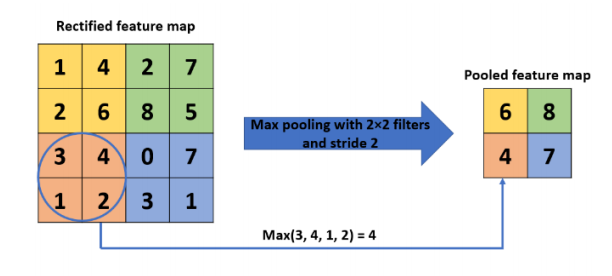}
  \caption{Example of a Max-Pooling Operation}
\end{figure}

The size of the output volume can be determined by this formula
if we have an activation map with dimensions $W \times W \times D$,
a pooling kernel with dimensions $F$ and a stride:

$$ W_{out} = \frac{W - F}{S} + 1 $$

This generates an output volume of $W_{out} \times W_{out} times D_{out}$.

The translation invariance of pooling makes it possible to
recognize objects wherever they appear in the frame regardless of their position.

\subsubsection{Fully connected layer}

As with regular FCNNs (Fully CNNs), neurons in this layer are fully
connected to neurons in the preceding and following layers.
Thus, it can be calculated as usual by a matrix multiplication
followed by a bias effect.
This layer enables mapping of inputs and outputs between representations.

\subsubsection{Activation layers}

Non-linear layers are often placed directly after the convolutional layer
to introduce nonlinearity to the activation map, due to the linear
nature of convolution and the non-linear nature of images.

\begin{enumerate}
  \item Sigmoid

        The sigmoid nonlinearity has the mathematical form
        $\sigma(\kappa) = 1 /  (1+ e^{-\kappa})$. This formula
        takes a real-valued number and ``squashes'' it between $0$
        and $1$. However, the gradient of the sigmoid is
        almost zero when the activation is at either tail.
        In backpropagation, if the local gradient becomes
        very small, it will effectively ``kill'' the gradient.
        Furthermore, if the sigmoid is always positive,
        it will produce either all positives or all negatives,
        resulting in a zig-zag trend in gradient updates for the weights.

  \item Tanh

        Tanh squashes a real-valued number between $-1$
        and $1$. The activation of sigmoid neurons saturates,
        but the output is zero-centered, unlike sigmoid neurons.

  \item ReLUs

        In the last few years, Rectified Linear Units (ReLUs)
        have been very popular. They are computed with the
        function $f(\kappa) = \max(0,\kappa)$.
        In other words, the activation threshold is
        simply set to  zero. With ReLUs, convergence
        is six times faster than the sigmoid and tanh non-linearities.

        The disadvantage of ReLUs is that they can be
        fragile during training.
        They can be updated by a large gradient in such a
        way that the neuron is never further updated.
        This can be addressed by setting an appropriate learning rate.

\end{enumerate}

\subsection{Cosine similarity for feature vector matching}

Humanode facial recognition system uses modified
ResNet architecture for facial feature extraction
and uses cosine similarity for matching.

Cosine similarity is a measurement that quantifies
the similarity between two or more vectors.
It is measured by the cosine of the angle between
vectors and determines whether two vectors are
pointing in roughly the same direction.
The vectors are typically non-zero and are
within an inner product space. It is described as
the division between the dot product of vectors and
the product of the Euclidean norms or magnitude of each vector:

$$similarity = \cos(\theta) = \frac{\mathbf{A}\cdot \mathbf{B}}{||\mathbf{A}|| ||\mathbf{B}||}
  = \frac{\sum_{i=1}^n A_i B_i}{\sqrt{\sum_{i=1}^n A_i^2} \sqrt{\sum_{i=1}^n B_i^2}} $$

and thus cosine similarities are constrained between $0$ and $1$.
The similarity measurement is a measure of the
cosine of the angle between the two non-zero vectors $\mathbf{A}$ and $\mathbf{B}$.

Assume the angle between the two vectors is $90$ degrees.
The cosine similarity will be zero in that case.
This indicates that the two vectors are orthogonal
or perpendicular to each other.
The angle between the two vectors $\mathbf{A}$ and $\mathbf{B}$
decreases as the cosine similarity measurement
approaches $1$. The image below illustrates this more clearly.

\begin{figure}[H]
  \centering
  \label{fig10}
  \includegraphics[width=0.75\linewidth, keepaspectratio]{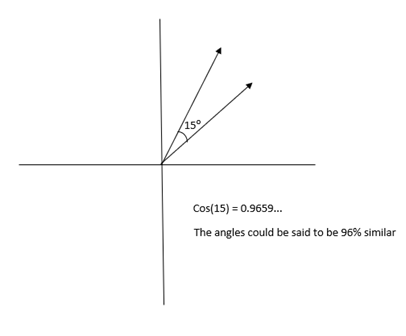}
  \caption{Two vectors with 96\% similarity based on the cosine of the angle between the vectors. }
\end{figure}

\begin{figure}[H]
  \centering
  \label{fig11}
  \includegraphics[width=0.75\linewidth, keepaspectratio]{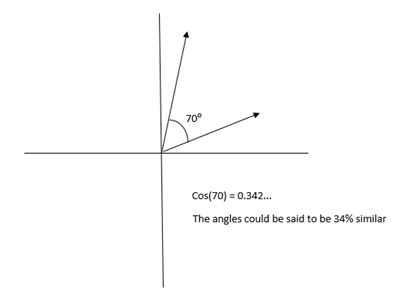}
  \caption{Two vectors with 34\% similarity based on the cosine of the angle between the vectors. }
\end{figure}

Humanode uses cosine similarity in the facial feature vector matching part.

\subsection{Active and Passive Liveness detection}

Enterprises use face recognition for onboarding, validating,
and approving customers due to its reliability and ease of use.
The demand for liveness detection is growing rapidly.
Liveness detection identifies presentation attacks like
photo or video spoofing, deepfakes, and 3D masks or models,
rather than matching the facial features.

This makes it much harder for an adversary to spoof an identity.
Facial recognition determines whether the person is
unique and the same whereas liveness detection determines
whether the person is a living human being.
Liveness detection confirms the presence of a user’s
identification credentials and that the user is
physically present, whether on a mobile phone, a
computer or tablet, or any camera-enabled device.

There are two methods in facial liveness detection: active and passive.

\textbf{Active liveness detection} asks the user to do
something to confirm that they are a live person.
A user would be normally asked to either change
the head position, nod, blink their eyes or follow
a mark on their device’s screen with their eyes.
Despite this, fraudsters can fool the active method
using a so-called presentation attack, also known as
the presentation attack detection (PAD) attack.
Scammers can use various gadgets or “artifacts” to fool
the system, some of which are remarkably low-tech.

The Humanode active liveness detection model asks
the user to turn their face left or right, blink their eyes,
or show emotions like happiness, anger, or surprise and
determines whether the user is fake or real depending on the result.

With \textbf{passive liveness detection}, the user is not
asked to do anything. This provides end-users with a
modernized and convenient experience.
It is an excellent method for determining whether
the user is present without any specific movement or gesture.
Passive methods use a single image,
which is examined for an array of multiple
characteristics to determine if a live person is present.

The Humanode passive liveness detection model
determines if a live person is present,
based on texture and local shape analysis,
distortion analysis, and edge analysis:

\begin{itemize}
  \item Texture and local shape analysis—analyze
        the input image from a textural analysis point
        of view by image quality assessment, characterization
        of printing artifacts, and differences in light reflection

  \item Distortion analysis—analyze the input image using
        an image distortion analysis feature vector that
        consists of four different features, specular reflection,
        blurriness, chromatic moment, and color diversity

  \item Edge analysis—analyze the edge of the input
        to find out whether the edge component is presented or not
\end{itemize}

\begin{figure}[H]
  \centering
  \label{fig12}
  \includegraphics[width=0.75\linewidth, keepaspectratio]{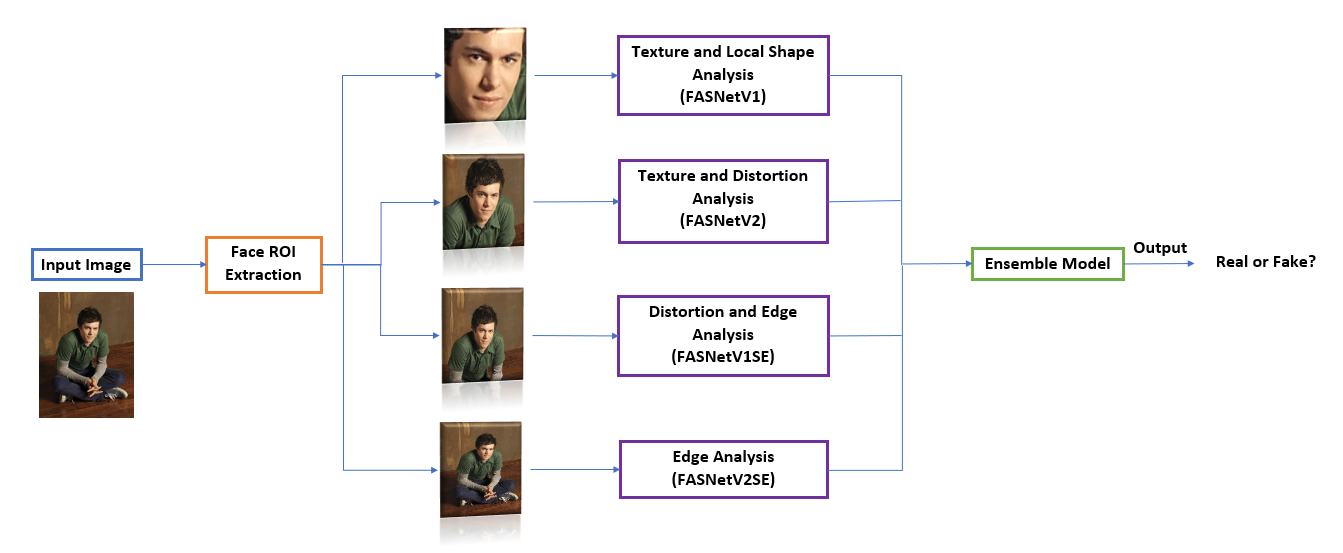}
  \caption{Analysis types in liveness detection. }
\end{figure}

While the active liveness detection process is going on,
passive liveness detection is performed in the background.

By combining the advantages of active and passive
liveness detection approaches, we made our liveness detection
system more secure.

\subsection{Merits and demerits of biometric identification}

The use of biometrics, the science of analyzing
physical or behavioral characteristics unique to each
individual to recognize their identity, has many benefits.
However, there are some risks associated with biometric authentication,
which are as follows.

\begin{table}[H]
  \centering
  \caption{Merits and demerits of biometric identification.}
  \label{table2}
  \resizebox{\textwidth}{!}{
    \begin{tabular}{|p{7cm}|p{6cm}|}
      \hline
      Merits & Demerits \\
      \hline
      \begin{itemize}
        \item \emph{High levels of security and accuracy} in contrast to passwords,
              as biometric data cannot be forgotten.

        \item \emph{Simplicity and convenience for the user} is a
              significant factor in the growing popularity of biometric authentication.

        \item \emph{Higher level of authenticity} for users prone to
              weak passwords that may be common to multiple users or easily shared.

        \item \emph{Affordability,} as biometric authentication is
              now possible in a wide range of common devices.

        \item \emph{Flexibility,} as users have their own security
              credentials with them, so they do not need to
              bother memorizing a complex password.

        \item \emph{Biometrics is trustable,} as reports from 2021
              claim that the younger generations trust biometric
              solutions more than others.

        \item Biometric solutions are \emph{time conserving.}
      \end{itemize}

             &

      \begin{itemize}
        \item Sometimes \emph{requires integration and/or
                additional hardware.}

        \item \emph{Delay,} as some biometric recognition methods
              may take more than the accepted time.

        \item \emph{Physical disability,} as some people are
              not fortunate enough to be able to
              participate in the enrollment process.

        \item The need to \emph{trust} your biometric
              provider that data are secure and private.
      \end{itemize}
      \\
      \hline
    \end{tabular}}
\end{table}

\subsection{Cryptobiometric search and matching operations}

When the user registers in the system, the executed
private neural network allows the feature vector
to be extracted from the user's face for the first time.
It is essential to safely store this vector to evaluate
the subsequent times that the user wants to authenticate
in the system. But this storage must be encrypted.
Moreover, to compare the new vector with the already
stored one, we cannot decrypt the data. For this, there
is the homomorphic encryption method.

Homomorphic encryption is nothing more than an encryption
algorithm with the additional characteristic that operations
can be defined so that they can be preserved by encryption.
In mathematics, the preservation of an operation is obtained
when we have an operation and a function between two spaces.
The function that goes from one space to the other is said
to preserve the operation if it is invariant under said operation.

Formally we say that the function $f$, from space $A$ to space $B$,
is an homomorphism if given two elements $a_1, a_2 \in A$ , then
the function $f$ has the following property:

$$f(a_1 + a_2) = f(a_1) + f(a_2)$$

This section will discuss a method used in neural networks
to evaluate the similarity between two feature vectors.
Then, we will define the homomorphic encryption method
that will allow us to store the encrypted feature vector
and perform the similarity operation without decrypting the vector.

\subsubsection{Cosine similarity encryption}

As mentioned above, one of the most efficient and
natural ways to find the similarity between two
feature vectors in neural networks is cosine similarity.
Let $c=(c_0, \ldots, c_r)$ and $b=(b_0, \ldots, b_n)$ two vectors in $$\mathbb{R}^n$$,
the cosine similarity between $a$ and $b$ is defined by the equation

$$cos(a,b) = \frac{a\cdot b}{||a|| ||b||}$$

where $$||a||$$ is the norm of the vector $a$ \cite{Kikuchi2008}.

From the equation above, if we calculate the internal
product between two vectors, we can determine if two vectors
are similar directly. In simple terms, the cosine similarity
of the angle of two vectors tells us whether two vectors
point in the same direction.

If in addition the vectors are normalized, then it is evident that:

$$cos(a,b) = \frac{a\cdot b}{||a|| ||b||} = a \cdot b = \sum_{i=1}^n a_i b_i$$

In the cryptobiometric authentication system, we must define
an encryption scheme that allows us to calculate the internal
product between two vectors, which will give us the similarity
between them. This calculation will be carried out on the
encrypted vectors without the need to decrypt them.

It is natural to look for a homomorphic encryption scheme
where the calculations to determine similarity are
performed in the encrypted space.

In a traditional encryption scheme, which only encrypts
the data to be sent, it would have to handle the private
keys with which the user encrypted the data, decrypt the vectors,
and then make the similarity calculation on clear data.
From a decentralized perspective, this traditional approach has a flaw,
as users' private keys are in an environment where
peers are by nature untrusted.
In a decentralized environment, there is no trusted third
party to handle the keys securely.

\subsubsection{Homomorphic encryption}

There are different proposals for encryption schemes
that preserve operations in a homomorphic manner
through the encryption function.
In particular, one of the most straightforward and most
efficient is encryption based on learning with errors (LWE) \cite{Gentry2013}.
In this section we will examine the mathematical
preliminaries of this cipher and the algorithms that compose it, namely:

\begin{itemize}
  \item Key generation
  \item Encryption
  \item Decryption
  \item Homomorphic operations
\end{itemize}

\textbf{\textit{Lattices}}

In group theory a lattice in $\mathbb{R}^n$ is an algebraic
subgroup of $\mathbb{R}^n$ that spans the vector space $\mathbb{R}^n$
with integer coefficients in its basis.

Formally, let $n\in \mathbb{N}$,$B \in \mathbb{R}^{n\times n}$ be a matrix,
and $$b_i \in \mathbb{R}^n$$ the $i$-th row of $B$ with $1\leq i \leq n$.
Then the linear combinations of $b_i$ are defined as

$$\mathcal{L} (B) = \sum_{i=1}^n m_i b_i$$

is a subgroup of $\mathbb{R}^n$, where $m_i \in\mathbb{Z}$.
If the $$b_i$$ are linearly independent,
we say that $\mathcal{L} (B)$ is a lattice in $\mathbb{R}^n$ of dimension $n$.

Lattice-based ciphers are some of the leading candidates for
post-quantum cryptographic algorithms.
If an efficient quantum computer is ever built,
a post-quantum encryption scheme can resist attacks.
In 1994, Shor theoretically demonstrated that a protocol
could be built on a quantum computer that would break in
polynomial time the problems on which most public-key
ciphers such as the RSA, Diffie–Hellman, or cryptosystems
of elliptic curves are based.

However, the computational complexity of the problem that
shapes cryptosystems based on lattices ensures their
quantum resistance. Furthermore, the LWE-based cryptosystem
can be completely homomorphic: it possesses homomorphism in
both operations of addition and multiplication. Which is
very useful for the calculation of the inner product, and
consequently for the similarity of the cosine \cite{Dinh2017}.

\subsubsection*{\emph{Construction of the ring-LWE scheme}}

Let us now see in detail how the ring-LWE encryption
scheme works and how the homomorphic operations are defined \cite{Abidin2014, BrakerskiVaikuntanathan2011}.

\paragraph{Setup parameters}

First of all, we need to define certain general parameters
to be used in the key generation algorithm:

\begin{itemize}
  \item set $n\in\mathbb{N}$, a degree parameter.

  \item let $q$ be a prime number, defining the ring $R_q = R/qR = \mathbb{F}_{q}[x]/f(x)$.
        This ring is the ciphertext space.

  \item take $t$ as an arbitrary integer, with $t < q$, defining the ring $R_t = R/tR = \mathbb{F}_{t}[x]/f(x)$.
        This ring is the plaintext space.

  \item define the standard deviation $\sigma$, as the parameter for the
        discrete Gaussian distribution $\chi = D_{Z^n, \sigma}$.
\end{itemize}

\paragraph{Key generation}

First, we sample random elements as follows:
\begin{itemize}
  \item sample $s$ from the Gaussian distribution $\chi$

  \item take a random $p_1\in R_q$ and the error $e$ sampled from $\chi$.
\end{itemize}

Then the public key is defined as
$pk=(p_0, p_1)$, where $p_0 = - (p_1 s + te)$, and the secret key is $sk=s$.

\paragraph{Encryption}

After encoding the plaintext $m$ as an element in $R_t$ and given the public key
$pk=(p_0, p_1)$, we sample $u, f, g$ from the distribution $\chi$ and compute

$$Enc(m,pk) = (c_0,c_1) = (p_0 u +tg +m, p_1 u + tf)$$

\paragraph{Decryption}

If $c = (c_0, \ldots, c_r)$ is a ciphertext and $sk=s$ the private key,
then the decryption is simply

$$Dec(c,sk) = [\widehat{m}]_q (mod t) \in R_t$$

where

$$\widehat{m}=\sum_{i=0}^r c_i s^i \in R_q$$

If we write the secret key vector $S$ as $S=(1, s, s^2, \ldots, s^r)$, then

$$Dec (c,sk)= [\langle c, S \rangle]_q (mod t)$$

\paragraph{Homomorphic operations}

Now, if we have two elements in the encrypted space,
$c=(c_0, \ldots,c_r)$, $c'=(c_0', \ldots,c_t')$
the homomorphic operations are given by

$$c + c' = (c_0 + c_0', \ldots,c_{max(r,t)} + c_{max(r,t)}')$$

$$x \ast c' = (\widehat{c}_0, \ldots, \widehat{c}_{r+t})$$

where

$$\sum_{i=0}^{r+t} \widehat{c}_i z^i= \left( \sum_{i=0}^r c_i z^i \right) \ast \left( \sum_{i=0}^t c_i' z^i \right)$$

\subsection{Extracting inner product from the encrypted value}

The cosine similarity operation requires, as we saw,
the calculation of the inner product in the encrypted space.
It is evident then that if we define accordingly a transformation
in the encrypted space, thanks to the homomorphic properties of the
encryption scheme we can extract the inner product as a constant
term from the encrypted result \cite{Chotard2017}.

Thus, let $P$, $Q$ be bit sequence representations of vectors
and $F$ a transformation onto the ring $R_q$ such that
$F(P) = \sum_{i=0}^{l-1} p_i 2^i $ and $F(Q) = \sum_{i=0}^{l-1} q_j 2^{n-j}$.

If we multiply $F(P)\ast F(Q)$, then

\begin{align*}
  F(P)\ast F(Q) & = \sum_{i=0}^{l-1} p_i q_i 2^n + \cdots \\
                & = \langle P, Q \rangle + \cdots
\end{align*}

Thus, if we encrypt $F(P)$ and $F(Q)$, thanks to the homomorphic properties
of the encryption scheme, we can extract the inner product as a constant
term from the encrypted result:

$$Enc (F(P)\ast F(Q)) = \langle P, Q \rangle + E(\cdots)$$

\subsection{ZKP for verifiable computation}

In our setup, a node does not trust any other node in the system.
This means that a node can be trusted to follow the protocol but may
not be trusted with the computation of the feature extraction and liveness
detection processes.

During the registration process, a node will extract a feature vector
from the face image and then send it to a peer node. The problem is how
does the peer node trust the feature vector? A node may or may not have
followed the feature extraction process as required. In this situation,
zero-knowledge–based verifiable computation comes to the rescue.

Verifiable computation is a technique to prove that the computation
process was followed correctly by an untrusted party. Let $y = f(x)$
be the result of computation on input $x$. The prover generates a proof of
computation, $\pi$, along with the result and sends $x,y,\pi$ to the verifier.
Using $x,y$ and verification keys, the verifier verifies the correctness
of the proof $\pi$.

\paragraph{Related work:}
\begin{enumerate}
  \item SafetyNet: Specialized interactive proof protocol for verifiable
        execution of a class of deep neural networks. It supports only quadratic
        activation functions, but in our neural network model, ReLU is necessary
        to achieve higher accuracy.

  \item  zkDT: Verifiable inference and accuracy schemes on decision trees.
        Decision trees are simple and quite different from neural network architecture.

  \item vCNN: verifiable inference scheme for neural networks with zero knowledge.
        It only optimizes convolution. The vCNN scheme uses mixing of QAP
        (Quadratic arithmetic program), QPP (quadratic polynomial program),
        and CP-SNARK for making a connection between QAP and QPP.
        QAP works at the arithmetic circuit level and is costly in terms of computation.

  \item ZEN: R1CS-friendly optimized ZK neural network inference scheme.
        Proposes an R1CS-friendly quantization technique. Uses an arithmetic-level
        circuit and Groth zero-knowledge–proof scheme.

  \item zkCNN: Interactive zero-knowledge–proof scheme for a CNN.
        Proposes a new sum-check protocol. Uses the GKR protocol.

\end{enumerate}

The vCNN, ZEN, and zkCNN procedures are most closely related to our scenario
but all of these reduce the computation program to arithmetic circuit level
and then use a Groth ZKP protocol for verification.

Any verifiable computation scheme utilizes the homomorphic property of the
underlying primitive for verification. Therefore, it can support computation
that involves either addition or multiplication. Since neural network computations
are often complex and non-linear, researchers often convert the program to the
arithmetic circuit level, which involves only addition and multiplication at
the bit level, and then use a zkSNARK-type proof. This is a more generalized
technique for any circuit. However, if the circuit involves only addition and
multiplication at integer level then there is no need to convert it to the arithmetic
circuit level.

Our idea is to break down the neural network model of feature extraction into
different layers and then prove the computation of individual layers separately.
There are four main layers: the convolution layer, Batch-normalization layer,
ReLU layer, and average pooling layer. Out of these, only the ReLU layer is
not in the form of addition and multiplication.

$$ReLU(x) = max(x,0)$$.

So, to make it compatible with our idea, we replaced the ReLU function with
the bit-decomposition of ReLU, which involves bit-level addition and multiplication.
After this, we used Verifiable Private Polynomial Evaluation (PIPE) where an
untrusted cloud server proves that the polynomial computation, $y = f(x)$,
is correct without revealing coefficients of the polynomial $f$.
We are aware of other similar schemes like Pinocchio,
PolyCommit by Kate et al., and other Garbled circuit-based schemes,
but PIPE is best suitable for our decentralized untrusted P2P network scenario.

Our scenario is similar but slightly different. We assume that the neural
network parameters are available with each node. That means coefficients
of the kernel in the convolution layer are available with each node.
For input $(x_1, \ldots, x_n)$ and kernel $(a_1, \ldots, a_m)$,
the output of convolution can be represented as:

$$y_j = \sum_i a_i x_{j+i}$$

In the PIPE scheme, $a_i$ is kept secret from the verifier and in
our scenario, $x_i$ (which represents input image) is kept secret
from the verifier. Moreover, in the PIPE scheme, both input and output
are available in plain form for the verifier.
However, we cannot reveal the input and outputs of the neural
network as well as the intermediate layers due to privacy concerns.
That means we had to modify the PIPE scheme in such a way that the
verifier can still verify the correctness of computation using encrypted
inputs and outputs.

Finally, here is what we have in a ZKP system for the feature vector extraction process.

\begin{figure}[H]
  \centering
  \label{fig13}
  \includegraphics[width=0.75\linewidth, keepaspectratio]{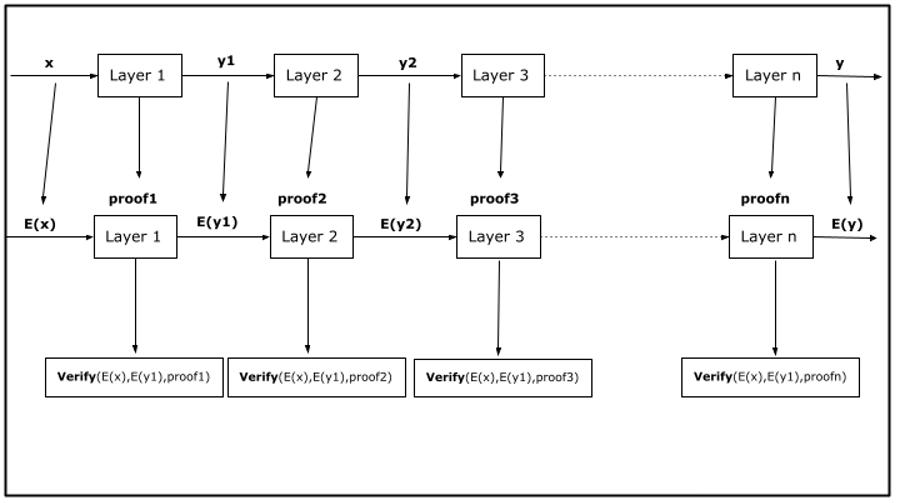}
  \caption{ZKP for the feature vector}
\end{figure}

\subsubsection{Humanode approach to ZKP}

\paragraph{Generalized problem:}

\textbf{Input:}  $(x_1, \ldots, x_n)$

\textbf{Computation:} $y=\sum_i a_i x_i$

Prover picks an input and performs the computation. Since the verifier does
not trust the prover, the prover needs to prove that the output $y$ is computed correctly.

\textbf{Requirement:} The coefficients of the computation,
$\{a_i\}$ are public and known to verifiers. The prover can’t disclose
$\{x_i\}$ and $y$  to the verifier due to privacy concerns.

To overcome this problem, we combined Feldman’s Verifiable Secret Sharing (VSS),
ElGamal Crypto system, and non-interactive ZK proof.

\subparagraph{Feldman’s Verifiable Secret Sharing scheme:}

Feldman’s VSS is a secret sharing scheme where each share is a point $(x.y)$ on
a secret polynomial $f$. In Feldman’s VSS, given a share $(a,b)$,
anybody can verify the validity of the share using some public value
corresponding to the secret polynomial $f$. This means anyone can check whether
$a = f(b)$ without knowing the coefficients of the polynomial $f$.

Let $f(x)=\sum_{i=0}^k a_i x^i$  be a $k$-degree polynomial with $a_i\in\mathbb{Z}_p^{\ast}$.

Let $G$ be a multiplicative group of a prime order $p$ and $g$
be a generator of $G$. For each $a_i$, set $h_i=g^{a_i}$.

Now make $g$ and $\{h_i\}$ public. Given a share $(a,b)$, one can check the
validity of the share by verifying the following equation:

$g^b = \prod_{i=0}^k h_i^{a_i} $

Note: There are two concerns here. First, the share $(a,b)$ is in plain form and hence,
if we use this as is, then we have to reveal input and output to the verifier in
our scenario. The second concern is that $h_i$ hides $a_i$ under the assumption
that it is difficult to solve for $a_i$ from $h_i$ under the discrete
logarithm assumption. However, if $a_i$ is a small value,
then it will be very easy to find $a_i$ from $h_i$.
In neural network computation, the values (input and weight parameters)
are always small and cannot hide it properly.

\subparagraph{Feldman’s VSS with encrypted input and output:}

\textbf{Input:}  $(x_1, \ldots, x_n)$

\textbf{Computation:} $y=\sum_i a_i x_i$

To hide input and output, we need to encrypt both in such a way
that we can perform some operation over encrypted value.
That means we have to use some homomorphic encryption scheme.

We use ElGamal encryption mainly because it is homomorphic with
respect to plaintext multiplication as well as scalar multiplication,
which suits our system perfectly.

ElGamal Key pair: = $(sk,pk) = (\alpha, h=g^{\alpha})$

Encrypt input:= $Enc(g^{x_i}) = (c_i,d_i) = (g^{r_i}, h^{r_i}g^{x_i})$

Encrypt Output:= $Enc(g^{y}) = (g^r, h^r g^y)$

Compute

$$C = \prod_i c_i^{a_i} = \prod_i g^{r_i a_i} = g^{\sum_i a_i r_i} = g^{r'}$$

where $r' = \sum_i a_i r_i$ and

$$D = \prod_i d_i^{a_i} = \prod_i (h^{r_i} g^{x_i})^{a_i} = (
  \prod_i h^{r_i a_i} )(\prod_i g^{x_i a_i}) = h^{r'} g^y$$

Finally, we have $(C,D) = (g^r,h^r,g^y)$, which is an ElGamal encryption of
$g^y$. So now, the prover needs to convince the verifier that $(C,D)$
computed from encrypted input is a valid ciphertext of $g^y$.
Here, we use the Non-Interactive ZK Proof (NIZKP) of $(C/g^r) = (D/h^r g^y)$

\subparagraph{Non-Interactive ZK Proof:}

If we generalize the above log equation, then we have $h_1=h_2$ for some
$g_1,h_1,g_2,h_2\in G$. In 1993, David Chaum and T. P. Pedersen proposed
the NIZKP to prove exactly this.

\textbf{NIZKP LogEq:}

Let $G$ be a multiplicative group of prime order $p$ and $H$ be a hash function.
Let the language $L$ be the set of all $(g_1,h_1,g_2,h_2) \in G^4$ where
$h_1 =h_2$ . The NIZKP $LogEq = (prove,verify)$ is as follows:

Prove $((g_1,h_1,g_2,h_2),w)$: Using the witness $w=h_1$, it picks a random
$r$ from $\mathbb{Z}_p^{\ast}$ and computes $A=g_1^r$, $B=g_2^r$ ,
$z=H(A,B)$  and $t=r+w.z$. It outputs proof $\pi=(A,B,t)$.

Verify $((g_1,h_1,g_2,h_2),\pi)$: Using $\pi=(A,B,t)$, it computes $z=H(A,B)$.
If

$$ g_1^t=A h_1^z \hbox{     and    }   g_2^t = B h_2^z $$

Then it outputs $1$, else it outputs $0$.

We achieve the ZKP system for an individual layer of our NN model by combining
Feldman’s VSS, ElGamal cryptosystem, and NIZKP LogEq properly.

\begin{itemize}
  \item Our ZKP system is unconditionally $ZK$-secure and $UNF$-secure
        under Random Oracle Model.

  \item Our ZKP system is also privacy-preserving under the DDH assumption
        in the Random Oracle Model.
\end{itemize}

\subsubsection{ElGamal Cryptosystem}

We generalize the input image as a higher-dimensional vector
$(x_1,x_2,x_3,\ldots,x_n)$. Similarly, we assume the output of each layer
is again a higher-dimensional vector $(y_1,y_2,\ldots,y_m)$.
For each layer, we encrypt its input and output using ElGamal encryption.

The ElGamal public-key encryption scheme is defined as follows:

\begin{itemize}
  \item $Gen(\lambda)$ - returns $pk=(G,p,g,h)$ and $sk=\alpha$ where
        $G$ is a multiplicative group of prime order $p$, $g\in G$ and $h=g^{\alpha}$.

  \item $Enc_{pk}(m)$ - returns $(c,d)=(g^r,pk^r.m)$ where $r$ is a randomly
        chosen integer between $1$ and $(p-1)$.

  \item $Dec_{sk}((c,d))$ - returns $m=\frac{d}{c^{sk}}$.
\end{itemize}

In our scheme, we use 1024-bit prime $p$ to achieve the recommended security.
Note that ElGamal encryption uses randomized encryption and is not deterministic.
That means if the same message is encrypted twice then both ciphertexts will be
different. Thus each transaction will be indistinguishable and preserve the
privacy of the user. Moreover, ElGamal encryption is homomorphic with respect
to plaintext multiplication and scalar multiplication.

$$Enc(m_1) \ast Enc(m_2) = Enc(m_1 m_2) $$

$$Enc(m)^{\alpha} = Enc(m \alpha)$$

\subsubsection{Zero-knowledge–proof system for liveness detection}

The result of liveness detection is proved by sending the output of the
detection algorithm. This output comes in the form of a yes or no,
a Boolean result.

In a centralized system, the algorithm runs in a controlled environment
where the central authority manages the input and output. When the user
is given the ability to run the liveness detection algorithm on their
own there is the risk of a malicious user tampering with the result of
the algorithm. Errors can also occur in the transmission of data or
local failures in executing the algorithm and obtaining the results.

The system's decentralization includes the need to prove that the
result is obtained through a correct execution of the algorithm.
That is why in Humanode, we have an algorithm to generate proof of
the correctness of each function of the liveness detection process.
In addition, there will be a verification algorithm for the said proof,
thus having a zero-knowledge–proof system suitable for decentralized
testing of the correct execution of liveness detection.

\subsubsection{Collective Authority}

One of the most critical problems to solve when defining encryption schemes
in decentralized environments is the handling of cryptographic keys, where
in addition, the calculations are performed and verified by peers through
multi-party computation.

In this sense, we will consider a subgroup of the Humanode network,
which we will call the Collective Authority, whose objective is to
generate the collective keys for homomorphic encryption and also verify
the calculations performed by each peer.

In simple terms, the Collective Authority works as a trusted third
party for key generation and verification but is also composed of several
peers within the network.

During the Setup process, the Collective Authority is the one who defines
generic parameters for the establishment of the cryptographic protocols \cite{UnLynx2017}.
The security that this Collective Authority provides us is that each peer
takes these generic parameters and locally generates its public and private keys,
as we saw above.

Each user keeps his private key secured locally but sends the public key to
the Collective Authority. After collecting the public keys from each user,
the Collective Authority constructs a collective public key and distributes
it back to all users \cite{UnLynx2017}.  This collective public key is the one used to encrypt
the feature vectors.

If a malicious user intercepts the public key in a traditional cryptosystem,
obtaining the private key is computationally challenging. In our case,
if the collective public key is intercepted, the perpetrator can't get
the private keys, as he must know which partial element belongs to which peer.
Thus we have an additional layer of security to the public-key cryptosystem, which we can
call a lattice-based decentralized public-key cryptosystem.

\subsection{Humanode's multimodal biometric approach}

The Biometric Identification Matrix was created by the Humanode core to
understand which of the existing biometric modalities are the most suitable
and superior and, therefore, to choose the proper ones for Humanode biometric
processing methods.

According to recent studies, \emph{there are three types of biometric measurements}
\cite{Kaur2014}:

\begin{itemize}
  \item Physiological measurement includes face recognition, finger or
        palm prints, hand geometry, vein pattern, eye (iris and retina),
        ear shape, DNA, etc.

  \item Behavioral measurement relating to human behavior
        that can vary over time and includes keystroke patterns,
        signature, and gait \cite{Jaiswal2011}.

  \item Some biometric traits act as both physiological and behavioral
        characteristics (e.g., brain waves or electroencephalography (EEG)).
        EEG depends on the head or skull shape and size, but it changes from
        time to time depending on circumstances and varies according to age.
\end{itemize}

In light of the latest developments, we propose a fourth measurement
—\emph{neurological}— as a part of both physiological (internal) and behavioral
measurements. We believe that neurosignatures, the technology of reading a human's
state of mind, i.e., signals that trigger a unique and distinct pattern of nerve
cell firing and chemical release that can be activated by appropriate stimuli,
should be developed and implemented in the Humanode as the most reliable and
secure way of biometric processing.

Until then, Humanode implements a multimodal biometric system of several
biometric modalities. Each biometric modality has its own merits and demerits.
It is laborious to make a direct comparison. Since the end of the 1990s,
when A. K. Jain, R. M. Bolle, and S. Pankanti conducted their comprehensive
research on all existing biometrics \cite{Jain1999}, seven significant
factors were identified to study and compare the biometric types: acceptability,
universality, uniqueness (distinctiveness), permanence, collectability,
performance, and resistance to circumvention—which are also known as
``\emph{the seven pillars of biometrics}'' \cite{JainRoss2004}.

Based on Jain et al.’s classification and recent all-encompassing surveys on
various biometric systems \cite{Weaver2000, Venkatesan2018}, cancelable systems \cite{Choudhury2018}, and unimodal, multimodal biometrics and
fusion techniques \cite{Raju2018}, we provide a
comparative study of different biometric modalities, and propose a 'Biometric
Identification Matrix', by studying and combining characteristics
revealed in the above-mentioned works and by adding factors we found
necessary to examine. Thus, we divided the ‘Performance’ category proposed
by Jain et al., which relates to the accuracy, speed, and robustness of
technology used, into two sub-categories ('Accuracy' and 'Processing Speed')
to study this space in more detail. To grasp how easy it is to collect
biometric data on a person, we decided to add the ‘Security’ category,
which refers to vulnerability to attack vectors, as paths or means by
which attackers can gain access to biometric data to deliver malicious actions.
The category ‘Hardware,’ which relates to the type of hardware, its prevalence,
and cost, was added to understand which devices are required to be used
nowadays and which are best to use in the network.

\begin{itemize}
  \item \textbf{Acceptability}
\end{itemize}

\emph{`Acceptability' relates to the relevant population’s willingness to
  use a certain modality of biometrics, their acceptance of the technology,
  and their readiness to have their biometrics trait captured and assessed.}

Complex and intrusive technologies have low levels of public acceptance.
Retina recognition is not socially acceptable, as it is not a very user-friendly
method because of the highly intrusive authentication process using retina
scanning \cite{Sadikoglu2016}. Electrophysiological methods (EEG, ECG) and
neurosignatures are not highly accepted nowadays, as they are intricate and
not yet well-known or fully developed.

An active liveness detection technology may be uncomfortable for the average
user if the trait acquisition method tends to be demanding or time-consuming.
Even in the absence of physical contact with sensors, many users still develop
a natural apathy for the entire liveness detection process, describing it as
over-intrusive (K. Okereafor \&Clement E. Onime, 2016).

\begin{itemize}
  \item \textbf{Collectability}
\end{itemize}

\emph{`Collectability' refers to the ease of data capturing, measuring,
  and processing, as well as reflecting how easy this biometric modality
  is for both the user and the personnel involved.}

Fingerprint and hand geometry recognition techniques are very easy to use.
Their template sizes are small and so matching is fast \cite{Jaiswal2011}.
Similarly, the advantages of face biometrics are that it is contactless,
and the acquisition process is simple. An advantage of all behavioral
recognition methods is the ease of acquisition as well.

\begin{itemize}
  \item \textbf{Permanence}
\end{itemize}

\emph{`Permanence' relates to long-term stability—how a modality varies over time.
  More specifically, a modality with 'high' permanence will be invariant over
  time with respect to the specific matching algorithm. }

Physiological measurements tend to be permanent, while behavioral
measurements are usually not stable over the long term. Such modalities
have a low or medium level of permanence.

The same person can sign in different ways, as a signature is affected by
physical conditions and feelings. Voice is not constant; it may change based
on an individual's emotion, sickness, or age \cite{Rabiner1993}.

Facial traits are persistent but may change and vary over time,
although heat generated by the tissues of the face has a measurable
repeatable pattern. It can be more stable than the facial structure
\cite{Hanmandlu2012}. Finger and palm prints and vein patterns tend
to remain constant. Hand geometry is more likely to be affected by disease,
weight loss/gain, injury. However, the results of hand geometry recognition
are not as much affected by skin moisture or textural changes depending on age.
Ear size changes over time \cite{Jaiswal2011, Abaza2013}. DNA is
highly permanent. The iris remains the same throughout life \cite{Kaur2014, Bowyer2008}.
However, diabetes and some other serious diseases can
cause alterations. Likewise, retinal patterns can change during medical
conditions like pregnancy, blood pressure, other ailments, etc. \cite{Kaur2014}.

\begin{itemize}
  \item \textbf{Universality}
\end{itemize}

\emph{`Universality' means that every person using a system may have the modality.}

Different biometric systems have their own limitations, likewise the modalities.
For example, some people have damaged or eliminated fingerprints, hand geometry
is efficient only for adults, etc. Biological/chemical, electrophysiological,
and neurological (in theory) biometrics measurement categories should have the
highest level of universality.

\begin{itemize}
  \item \textbf{Uniqueness}
\end{itemize}

\emph{`Uniqueness' relates to characteristics that should be sufficiently
  different for individuals such that they can be distinguished from one another.}

Every person has a unique walking style as well as writing style and
hence a person has his own gate and signature. Voice recognition technology
identifies the distinct vocal characteristic of the individual. Even so,
human behavior is not as unique as physiological patterns.

Finger and palm prints are extremely distinctive. The blood vessels underneath
the skin are also unique from person to person. The iris is highly unique and
rich in texture. Moreover, the textures of both eyes are different from each other.
Each person has a unique body odor and such chemical agents of human body odor
can be extracted from the pores to recognize a person \cite{Shu2014}.
Although at one time, neuroscientists thought brain activity was pretty much
the same from one person to another \cite{Finn2015, Finn2019, Demertzi2019},
people display a distinct `brain signature' when they are processing information,
similar to fingerprints.

Nevertheless, even physical modalities have limitations.
Thus, faces seem to be unique, however, in the case of twins,
distinctiveness is not guaranteed. DNA itself is unique for each individual,
except identical twins, therefore, it achieves high accuracy. However,
retina recognition is highly reliable, since no two people have the same
retinal pattern and even identical twins have distinct patterns. We assume
that neurosignatures are to be one of the premier biometric technologies
because of the unique nature of human thoughts, memories, and other mental conditions.

\begin{itemize}
  \item \textbf{Accuracy}
\end{itemize}

\emph{`Accuracy' is a part of the ‘Performance’ category. It describes how well
  a biometric modality can tell individuals apart. This is partially determined
  by the amount of information gathered as well as the quality of the neural
  network resulting in higher or lower false acceptance and false rejection rates.}

Two-dimensional (2D) facial recognition may give inaccurate results,
as facial features tend to change over time due to expression, and other
external factors. Also, it is highly dependent on lighting for correct input.
Thermograms, which are easy to obtain and process, are invariant to
illumination and work more accurately even in dim light, are far better.

Three-dimensional (3D) face recognition has the potential to
achieve greater accuracy than its 2D counterpart by measuring the
geometry of facial features. It avoids such pitfalls of 2D face recognition as
lighting, makeup, etc. It is worth noting that 3D face recognition with
liveness detection is considered the best in accuracy.

Palm prints show a higher level of accuracy than fingerprints.
Considering the number of minutiae points of all five fingers, the palm
print has more minutiae points to help make comparisons during the matching
process compared to fingerprints alone \cite{Kong2009}.

The iris provides a high degree of accuracy (iris patterns match for
1 in 10 billion people; \cite{Daugman2004}), but still can be affected by
wearing glasses or contact lenses. Similarly, retinal recognition is a
highly accurate technology; however, diseases such as cataracts, glaucoma,
diabetes, etc. may affect the results.

\begin{itemize}
  \item \textbf{Security}
\end{itemize}

\emph{`Security' refers to vulnerability to attack vectors, i.e., paths or
  means by which attackers can gain access to users’ biometric data to deliver
  malicious actions.}

Vascular biometrics rank as the safest because of the many benefits it inherently
offers, it is simple and contact-free as well as resilient to presentation attacks.
This applies to both hand and eye vein recognition. The vein pattern is not visible
and cannot be easily collected like facial features, fingerprints, voice, or DNA,
which stay exposed and can be collected without a person’s consent.

However, face recognition offers appropriate security if the biometric system
employs anti-spoofing and liveness detection so that an impostor may not gain
access with presentation attacks. 3D templates and the requirement of blinking
eyes or smiling for a successful face scan are some of the techniques that
improve the security of face recognition.

\begin{itemize}
  \item \textbf{Processing Speed}
\end{itemize}

\emph{`Processing Speed' is part of the ‘Performance’ category. It is related
  to the time it takes a biometric technology to identify an individual. }

As different modalities have different computation requirements,
the processing power of the systems used varies. Fingerprints and face
recognition are still the fastest in terms of the identification process.
The time used by vein recognition systems is also very impressive and reliable,
in terms of comparison of the recorded database to that of the current data.
Currently, the time which is taken to verify each individual is shorter than
other methods (average is 0.5 seconds; \cite{Le2011}). Iris and retina
recognition have a small template size, hence a promising processing
speed (2 to 5 seconds). Ear shape recognition techniques demonstrate
faster identification results, thanks to reduced processing time.
The more complicated the procedure, the longer it takes. Behavioral
modality identification can be processed fast. Signature, voice, and
lip motion recognition take a few seconds. The EEG and ECG processes differ.
Acquisition of a DNA sample requires a long procedure to return results
\cite{Bhable2015}.

\begin{itemize}
  \item \textbf{Circumvention}
\end{itemize}

\emph{`Circumvention' relates to an act of cheating; thus, the identifying
  characteristic used must be hard to deceive and imitate using an artifact
  or a substitute.}

Nearly every modality may become an easy subject for forgers.
Signatures can be effortlessly mimicked by professional attackers;
voices can be simply spoofed. Fingerprints are easily deceived through
artificial fingers made of wax, gelatin, or clay. Iris-based systems
can be attacked with fake irises printed on paper or wearable plastic
lenses, while face-based systems without five levels of liveness
detection can be fooled by sophisticated 3D masks \cite{Le2011}.
Even vein patterns can be imitated by developing a hand substitute.

Having said that our DNA is left everywhere, and has no inherent liveness,
it is believed to be the most difficult characteristic to dupe, as the
DNA of each person is unique \cite{Le2011}. Brain activity and heartbeat
patterns are also hard to emulate.

\begin{itemize}
  \item \textbf{Hardware}
\end{itemize}

\emph{`Hardware' category refers to the type and cost of hardware required
  to use the type of biometric. }

Nowadays, there is no need for extra new devices if you have a
smartphone for biometric recognition. Facial recognition and fingerprints
are common features of smartphones. For lip motion recognition, existing
image capturing devices, i.e., cameras, can be used. Thermograms need specialized
sensor cameras. Voice recognition is also easy to implement on smartphones or
any audio device. Hand vein recognition has a low cost in terms of installation
and equipment. Nowadays, mobile apps for vascular biometric recognition are
integrated using the palm vein modality \cite{Garcia2020}.
For eye vein identification, smartphones are currently in development, while retina
recognition is still an expensive technology, i.e., a high equipment cost.
Keystrokes need no special hardware or new sensors, and low-cost identification
is fast and secure. Image-based smartphone application prototypes for ear
biometrics are in development \cite{Bargal2015, Abate2016}, as well as mobile apps
with digital signatures \cite{Rahmawati2017}.

In the meantime, electroencephalograms are needed for EEG, and
electrocardiograms for ECG. Brain-computer interfaces (BCI) are needed for
neurosignatures. Special expensive equipment and hardware are needed for
DNA-matching procedures.

\subsubsection{Neurosignatures and other emerging biometric modalities}

We assume that a combination of the above-mentioned biometrics methods
(and even multimodal biometrics) are not one hundred percent safe/secure.
In the future, we plan to expand the system with this multimodal scheme,
making neurosignatures one of the main methods of Humanode user
identification/verification.

Other emerging modalities to research and to possibly implement in Humanode’s
verification system are as follows \cite{Goudelis2009}: smile recognition,
thermal palm recognition, hand/finger knuckle, magnetic fingerprints/smart
magnet, nail ID, eye movement, skin spectroscopy, body salinity, otoacoustic
emission recognition (OAE), mouse dynamics, palate, dental biometrics, and
cognitive biometrics.

\subsubsection{Biometric Identification Matrix}

\begin{table}[H]
  \centering
  \caption{`Biometric Identification Matrix': Biometrics Techniques Comparison.}
  \label{table3}
  \resizebox{\textwidth}{!}{
    \begin{tabular}{|p{2.5cm}|p{3cm}|c|c|c|c|c|c|c|c|c|c|}
      \hline
                                                                &                                    & \multicolumn{10}{c|}{Characteristics/Factors (L—Low, M—Medium, H—High) }                                                                 \\
      \hline
      Measurements                                              & Biometric Modality                 & Acceptability                                                            & Collectability & Permanence       &
      Universality                                              & Uniqueness                         & Accuracy                                                                 & Security       & Processing Speed &
      Circumvention                                             & Hardware                                                                                                                                                                      \\
      \hline
      \multirow{13}{*}{\parbox{2cm}{External  Physiological}}   & 2D Facial Recognition              & H                                                                        & H              & M                &
      H                                                         & M                                  & M                                                                        & M              & H                & H & H                     \\
      \cline{2-12}
                                                                & 3D Facial Recognition              & H                                                                        & H              & M                & H & M & H & H & H & L & H \\
      \cline{2-12}
                                                                & Facial Thermography Recognition    & H                                                                        & H              & M                & H & H & H & H & H & L & L \\
      \cline{2-12}
                                                                & Fingerprint                        & H                                                                        & M              & H                & M & M & L & L & H & H & H \\
      \cline{2-12}
                                                                & Palm Print / Footprint Recognition & M                                                                        & M              & H                & M & M & M & L & H & M & M \\
      \cline{2-12}
                                                                & Finger/Hand Geometry Recognition   & M                                                                        & H              & L                & H & M & M & L & H & M & L \\
      \cline{2-12}
                                                                & Finger/Hand Vein Recognition       & M                                                                        & M              & M                & M & M & H & H & H & L & L \\
      \cline{2-12}
                                                                & Iris Recognition                   & L                                                                        & H              & H                & H & H & H & H & M & L & M \\
      \cline{2-12}
                                                                & Retina Recogntion                  & L                                                                        & M              & H                & H & H & H & H & M & L & L \\
      \cline{2-12}
                                                                & Eye Vein Recognition               & L                                                                        & L              & H                & H & H & H & H & M & L & M \\
      \cline{2-12}
                                                                & Ear Shape                          & H                                                                        & M              & H                & M & L & L & L & M & M & M \\
      \cline{2-12}
                                                                & Body Odor                          & M                                                                        & L              & H                & H & L & L & M & M & L & L \\
      \cline{2-12}
                                                                & DNA Matching                       & L                                                                        & L              & H                & H & H & H & L & L & L & L \\
      \hline
      \multirow{2}{*}{\parbox{2cm}{Physiological + Behavioral}} & Brain Activity (EEG)               &
      L                                                         & L                                  & M                                                                        & H              & H                & H & H & L & L & L         \\
      \cline{2-12}
                                                                & Electrocardiography (ECG)          & L                                                                        & L              & M                & H & M & H & H & L & L & L \\
      \hline
      Internal Physiological + Behavioral                       & Neurosignatures                    & L                                                                        & L              & H                & H &
      H                                                         & H                                  & H                                                                        & H              & L                & L                         \\
      \hline
      \multirow{5}{*}{Behaviormetrics}                          & Gait                               & M                                                                        & H              & M                & M & M & L & L & M & M & L \\
      \cline{2-12}
                                                                & Keystroke Dynamics                 & H                                                                        & H              & L                & L & L & L & M & M & M & M \\
      \cline{2-12}
                                                                & Lip Motion Recognition             & H                                                                        & H              & M                & H & L & L & M & H & L & H \\
      \cline{2-12}
                                                                & Signature Recognition              & H                                                                        & H              & L                & L & L & M & L & H & H & M \\
      \cline{2-12}
                                                                & Voice Recognition                  & H                                                                        & H              & L                & M & M & M & L & H & H & H \\
      \hline
    \end{tabular}}
\end{table}

The different biometrics techniques are discussed. The advantages and
disadvantages associated with each of them are listed in Table \ref{table4}.

\begin{center}
  \begin{longtable}{|p{2.3cm}|p{3cm}|p{4.5cm}|p{5cm}|}
    \caption{`Biometric Identification Matrix': Biometrics Techniques Pros and Cons.\label{table4} }                                                                                                                                                                 \\
    \hline
    Measurements                                                       & Biometric Modality                                                                                          & Pros                                                                   & Cons \\
    \hline
    \multirow{13}{*}{\parbox{2cm}{External Physiological}}             & Facial Recognition                                                                                          &
    non-intrusive, fast, easy to set up, no additional hardware needed &
    face recognition systems are vulnerable to manipulation and impostor
    attacks, errors: lighting, age, glasses, hair                                                                                                                                                                                                                    \\
    \cline{2-4}
                                                                       & 3D Facial Recognition                                                                                       & easy, fast, and accurate                                               &
    Potential attacks of lifelike dolls, realistic 3D masks that may bypass the system                                                                                                                                                                               \\
    \cline{2-4}
                                                                       & Facial Thermography Recognition                                                                             &
    non-intrusive, non-invasive, more stable than the facial structure,
    does not depend on external illumination                           &
    illnesses, high cost of implementation, more expensive                                                                                                                                                                                                           \\
    \cline{2-4}
                                                                       & Fingerprint                                                                                                 &
    inexpensive, socially acceptable, easy to set up and easy to collect,
    ability to enroll multiple fingers                                 &
    easily deceived through artificial finger made of wax, cuts, scars,
    or absence of finger can
    produce obstacle for the recognition process                                                                                                                                                                                                                     \\
    \cline{2-4}
                                                                       & Palm Print / Footprint Recognition                                                                          &
    has more minutiae points to make comparisons during the matching
    process compared to fingerprint, does not pose high-security threats
                                                                       & injuries, dryness, dirt, age, not suitable for high-security apps                                                                                                                           \\
    \cline{2-4}
                                                                       & Finger/Hand Geometry Recognition                                                                            &
    easy to use, simple and fast, can withstand harsh environmental
    conditions, not affected by surface condition of the skin
                                                                       & requires training for the users, needs a large space or
    sensor to acquire the hand geometry, and is not distinctive enough
    to distinguish over a large database, errors: diseases, weight loss/gain, injury, age                                                                                                                                                                            \\
    \cline{2-4}
                                                                       & Finger/Hand Vein Recognition                                                                                &
    the vein patterns tend to remain constant over a long period of time
                                                                       & visibility depends on the factors like age, mole, physical activity, thickness of the skin, etc.                                                                                            \\
    \cline{2-4}
                                                                       & Iris Recognition                                                                                            &
    iris remains stable for years, well protected from damage, possible from a distance
                                                                       & can be affected by age and eye diseases that
    deteriorate transparency of cornea, errors:
    reflection, poor lighting, eyelids, eyelashes, contact lenses, glasses, etc.                                                                                                                                                                                     \\
    \cline{2-4}
                                                                       & Retina Recognition                                                                                          &
    one of the most secure and extremely accurate methods              &
    expensive, special equipment is required, highly invasive,
    not socially acceptable, the pattern changes during medical
    conditions like pregnancy, blood pressure, other ailments, etc.                                                                                                                                                                                                  \\
    \cline{2-4}
                                                                       & Eye Vein Recognition                                                                                        & long-term stability,
    the technology works even with glasses or contact lenses
                                                                       & quality of images is affected by
    numerous factors such as body temperature and heat                                                                                                                                                                                                               \\
    \cline{2-4}
                                                                       & Ear Shape                                                                                                   &
    can use with existing cameras and image capture devices,
    does not require close proximity                                   &
    errors in recognition as the images are not ideal, unclear recognition due
    to the effect of hair, hats, and earrings, not believed to be very distinctive                                                                                                                                                                                   \\
    \cline{2-4}
                                                                       & Body Odors                                                                                                  & identification is possible
    by a mixture of characteristic odors and recognizing the mixture’s components
                                                                       & artificial noses are not comfortable, distinctiveness is reduced by
    deodorants and perfume                                                                                                                                                                                                                                           \\
    \cline{2-4}
                                                                       & DNA Matching                                                                                                & provides high accuracy, does not suffer from system performance issues
                                                                       & complex method, requiring a physical sample that has to be stored with appropriate environmental conditions                                                                                 \\
    \hline
    \multirow{2}{*}{\parbox{2cm}{Physiological + Behavioral}}          & Brain Activity (EEG)
                                                                       & high security and accuracy
                                                                       & time-consuming and expensive process, brain signals for the
    specific task might change during different circumstances and a
    person can change his/her own brainwave pattern                                                                                                                                                                                                                  \\
    \cline{2-4}
                                                                       & Electrocardiography (ECG)                                                                                   & high security and accuracy
                                                                       & time-consuming and complex process                                                                                                                                                          \\
    \hline
    Internal Physiological + Behavioral                                & Neurosignatures
                                                                       & the most secure type, easy to use one's mental state, conscious state,
    or simply motor signals from the cortex
                                                                       & highly intrusive, no technology yet                                                                                                                                                         \\
    \hline
    \multirow{5}{*}{Behaviometric}                                     & Gait
                                                                       & unobtrusive method, easy to set up, video footage from existing surveillance
    cameras can be used
                                                                       & injuries, low reliability of results, computationally expensive since it
    requires more computations                                                                                                                                                                                                                                       \\
    \cline{2-4}
                                                                       & Keystroke Dynamics                                                                                          & works in the background, needs no special
    hardware, low cost
                                                                       & hand injury, tiredness, gap in days, change of keyboard etc.
    can change the typing rhythm                                                                                                                                                                                                                                     \\
    \cline{2-4}
                                                                       & Lip Motion Recognition                                                                                      &
    fixes shortcomings associated with classic biometric methods,
    easy to set up, interaction of a user is not necessary and can be used
    without the knowledge of the user                                  &
    still in its infancy, the relevant information may not be
    acquired from the specific facial attributes                                                                                                                                                                                                                     \\
    \cline{2-4}
                                                                       & Signature Recognition                                                                                       & wide acceptance in public, non-invasive in nature,
    easy to restore the template if it is stolen.
                                                                       & changing or evolving signatures, excludes people who are illiterate,
    and people who are not able to write their signature                                                                                                                                                                                                             \\
    \cline{2-4}
                                                                       & Voice Recognition                                                                                           & reliable and easy to use                                               &
    prone to spoofing attacks, a massive amount of storage is needed,
    technology is highly affected by the background noise                                                                                                                                                                                                            \\
    \hline
  \end{longtable}
\end{center}

\subsubsection{Humanode Biometric Modalities Score}

We assigned each factor its own value depending on its effectiveness
in terms of enrollment of new human nodes to the network:

\begin{itemize}
  \item Acceptability (6)

  \item Collectability (6)

  \item Permanence (5)

  \item Universality (5)

  \item Uniqueness (10)

  \item Accuracy (8)

  \item Security (10)

  \item Processing Speed (3)

  \item Circumvention (10)

  \item Hardware (8)
\end{itemize}

Thus, we assume that the most significant for the network are `Uniqueness'
and `Security' of the biometric modality, `Accuracy' of the biometric method,
a low level of `Circumvention', and the `Hardware' type used.

To evaluate every above-mentioned biometrics modality technique,
we proposed the `Humanode Biometric Modalities Score',
based on the `Biometric Identification Matrix' analyzed in Table \ref{table4}.

The study revealed that the 3D facial recognition technique has the highest
score (198), with facial thermography recognition (192) and iris recognition (190)
not far behind. Retina recognition (176) and eye vein recognition (178) also got
quite high scores, as well as neurosignatures (173), which are not so highly
scored because they are not yet fully developed or widely adopted.

\begin{table}[H]
  \centering
  \caption{`Biometric Identification Matrix': Modalities Scores.}
  \label{table5}
  \resizebox{\textwidth}{!}{
    \begin{tabular}{|p{2.2cm}|c|c|c|c|c|c|c|c|c|c|p{2cm}|}
      \hline
      \multirow{2}{*}{\parbox{2cm}{Biometric Modality}} & \multicolumn{10}{c|}{Characteristics/Factors (1—Low, 2—Medium, 3—High) }
                                                        & \multirow{2}{*}{\parbox{2cm}{Modalities Score}}                                                                                                                                       \\
      \cline{2-11}
                                                        & Acceptability (6)                                                        & Collectability (6) & Permanence (5)       & Universality (5)   & Uniqueness (10)
                                                        & Accuracy (8)                                                             & Security (10)      & Processing Speed (3) & Circumvention (10) & Hardware (8)    &                         \\
      \hline
      Facial Recognition                                & 3                                                                        & 3                  & 2                    & 3                  & 2               & 2 & 2 & 3 & 1 & 3 & 160 \\
      \hline
      3D Facial Recognition                             & 3                                                                        & 3                  & 2                    & 3                  & 2               & 3 & 3 & 3 & 3 & 3 & 198 \\
      \hline
      Facial Thermography Recognition                   & 3                                                                        & 3                  & 2                    & 3                  & 3               & 3 & 3 & 3 & 3 & 1 & 192 \\
      \hline
      Fingerprint                                       & 3                                                                        & 2                  & 3                    & 2                  & 2               & 1 & 1 & 3 & 1 & 3 & 136 \\
      \hline
      Palm Print / Footprint Recognition                & 2                                                                        & 2                  & 3                    & 2                  & 2               & 2 & 1 & 3 & 2 & 2 & 140 \\
      \hline
      Finger/Hand Geometry Recognition                  & 2                                                                        & 3                  & 1                    & 3                  & 2               & 2 & 1 & 3 & 2 & 1 & 133 \\
      \hline
      Finger/Hand Vein Recognition                      & 2                                                                        & 2                  & 2                    & 2                  & 2               & 3 & 3 & 3 & 3 & 2 & 170 \\
      \hline
      Iris Recognition                                  & 1                                                                        & 3                  & 3                    & 3                  & 3               & 3 & 3 & 2 & 3 & 2 & 190 \\
      \hline
      Retina Recognition                                & 1                                                                        & 2                  & 3                    & 3                  & 3               & 3 & 3 & 2 & 3 & 1 & 176 \\
      \hline
      Eye Vein Recognition                              & 1                                                                        & 1                  & 3                    & 3                  & 3               & 3 & 3 & 2 & 3 & 2 & 178 \\
      \hline
      Ear Shape                                         & 3                                                                        & 2                  & 3                    & 2                  & 1               & 1 & 1 & 2 & 2 & 2 & 125 \\
      \hline
      Body Odor                                         & 2                                                                        & 1                  & 3                    & 3                  & 1               & 1 & 2 & 2 & 3 & 1 & 130 \\
      \hline
      DNA Matching                                      & 1                                                                        & 1                  & 3                    & 3                  & 3               & 3 & 1 & 1 & 3 & 1 & 147 \\
      \hline
      Brain Activity (EEG)                              & 1                                                                        & 1                  & 2                    & 3                  & 3               & 3 & 3 & 1 & 3 & 1 & 146 \\
      \hline
      Electrocardiography (ECG)                         & 1                                                                        & 1                  & 2                    & 3                  & 2               & 3 & 3 & 1 & 3 & 1 & 152 \\
      \hline
      Neurosignatures                                   & 1                                                                        & 1                  & 3                    & 3                  & 3               & 3 & 3 & 3 & 3 & 1 & 173 \\
      \hline
      Gait                                              & 2                                                                        & 3                  & 2                    & 2                  & 2               & 1 & 1 & 2 & 2 & 1 & 122 \\
      \hline
      Keystroke Dynamics                                & 3                                                                        & 3                  & 1                    & 1                  & 1               & 1 & 2 & 2 & 2 & 2 & 126 \\
      \hline
      Lip Motion Recognition                            & 3                                                                        & 3                  & 2                    & 3                  & 1               & 1 & 2 & 3 & 3 & 3 & 162 \\
      \hline
      Signature Recognition                             & 3                                                                        & 3                  & 1                    & 1                  & 1               & 2 & 1 & 3 & 1 & 2 & 117 \\
      \hline
      Voice Recognition                                 & 3                                                                        & 3                  & 1                    & 2                  & 2               & 2 & 1 & 3 & 1 & 3 & 131 \\
      \hline
    \end{tabular}}
\end{table}

* When calculated, we swapped the levels (numbers) for the ‘Circumvention’
factor so that it could be correlated with other factors, since a ‘High’ level
of circumvention means it is easy to imitate the body part, the modality, by
using an artifact or substitute, while a ‘Low’ level of circumvention means
this is practically impossible to do. In our model ‘Low’ is assin a value of
3 while ‘High’ is assigned the value -1.

\begin{figure}[H]
  \centering
  \label{fig14}
  \includegraphics[width=0.75\linewidth, keepaspectratio]{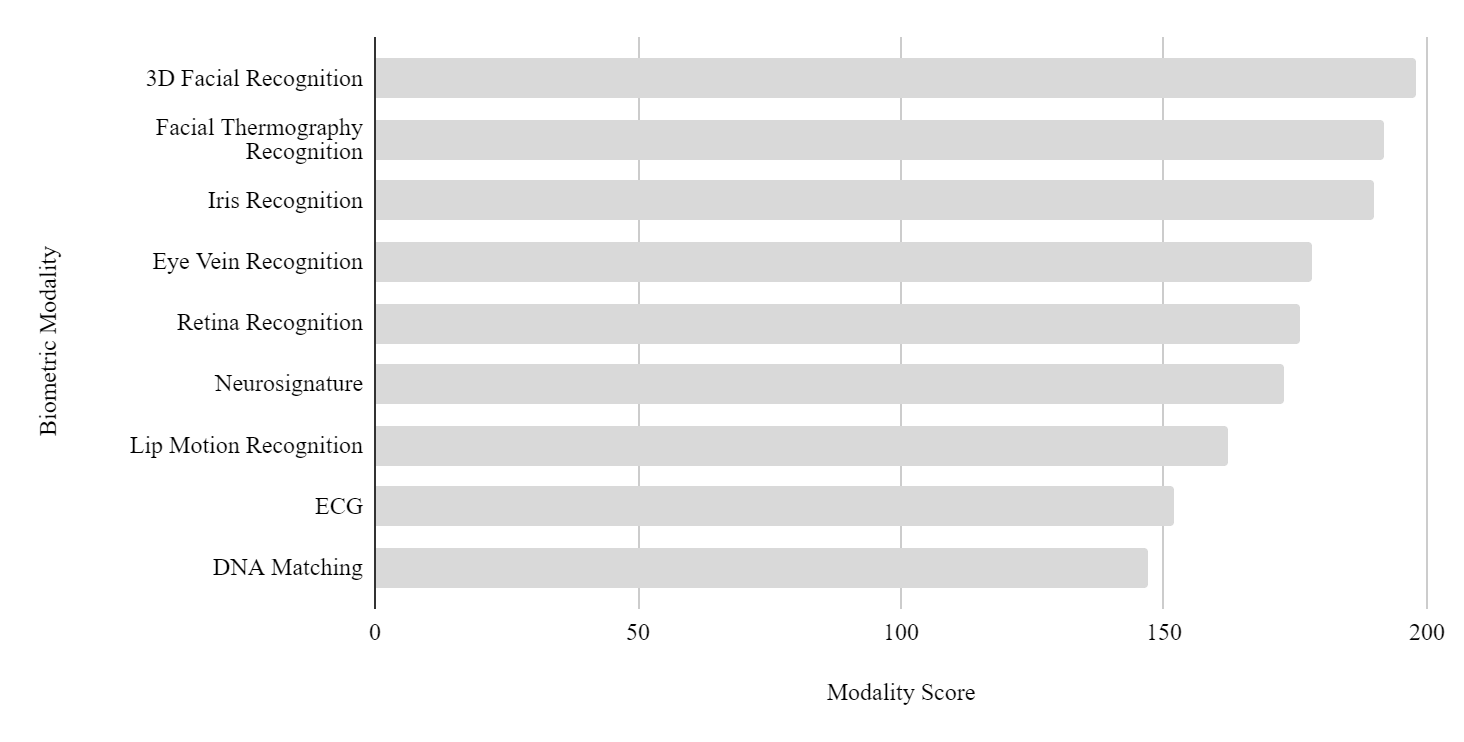}
  \caption{`Biometric Identification Matrix': Modalities Scores }
\end{figure}

To create a human node, only those modalities will be used that have score
points above the median value $(>147)$, i.e., 2D facial recognition, 3D facial
recognition, facial thermography recognition, iris, retina, finger/hand vein
recognition, eye vein recognition, ECG, DNA matching, and neurosignatures (in the future).

Due to the possible development of methods of attack on the current
biometric security setup in the future, the Humanode network will require
human nodes to provide additional biometric data during network upgrades.
For instance, once iris verification is shown to be secure on smartphone
devices, it will be added as an additional minimum requirement to deploy
a node. While Samsung already has made attempts \cite{Samsung} to deploy consumer-scale
iris recognition into its smartphones, its quality and security levels
are quite low compared to specialized hardware.

On top of this, in order to increase the cost of possible attacks on
biometrics, the Humanode network requires high standards for the
multimodal biometric system used for granting permission to launch a
human node. Starting only with 3D facial recognition and liveness detection,
later on one will have to go through multimodal biometric processing.

Also, the ability to create several wallets and to choose their types
in the system will be correlated with the biometric modalities selected.
For example, to create a high-value wallet, a more secure and complex
verification technique should be chosen, and vice versa.

\subsection{Types of attacks on biometric systems and their solutions }

Currently, there are eight possible attacks against biometric systems,
which are discussed below.

\begin{figure}[H]
  \centering
  \label{fig15}
  \includegraphics[width=0.75\linewidth, keepaspectratio]{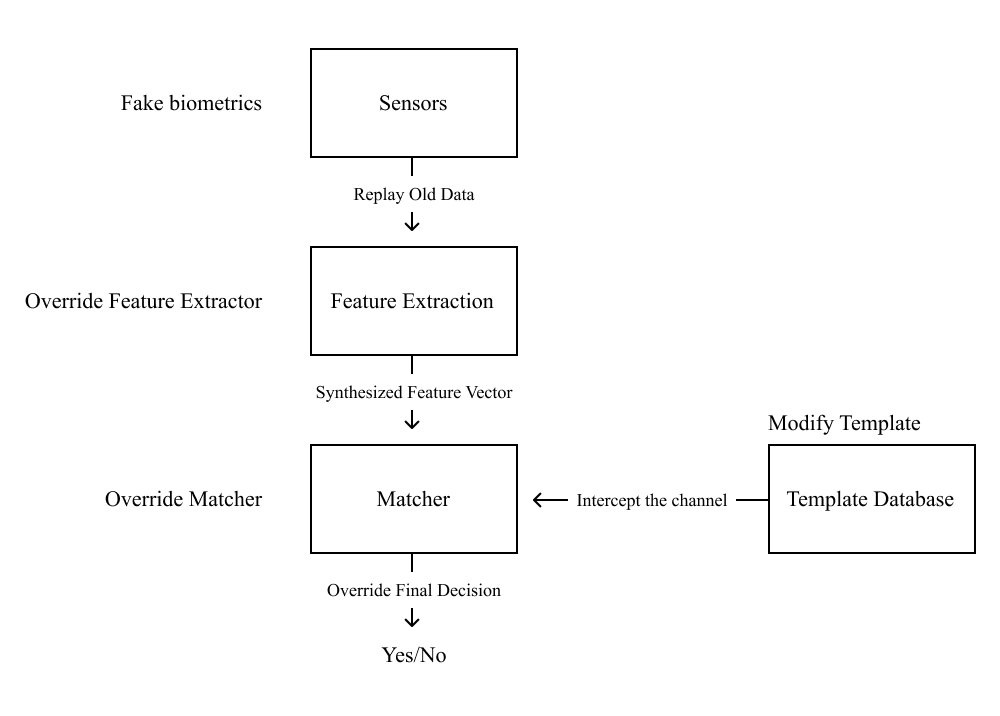}
  \caption{Possible attacks on biometric verification systems.}
\end{figure}

\subsubsection{Attack on the sensor}

Attackers can present fake biometrics to attempt to fool sensors
\cite{Jain2008}. For example, someone can make a fake hand with fake vein
patterns, or a finger with a fake wax fingerprint; wear special-made lenses
to bypass the iris scanner; and create images of a legitimate user to bypass
the face recognition system, etc. The possible solutions for this type of attack
are multimodal biometrics, liveness detection, as well as soft biometrics
\cite{Kamaldeep2011}.

\emph{Multimodal biometrics} is the main technique to prevent attacks and
make the biometric system more secure. Multimodal biometrics refers to methods
in which several biometric features are considered for enrollment and authentication.
When multiple biometric characteristics are used, it becomes difficult for an attacker
to gain access to all of them.

Humanode utilizes multimodal biometrics. The network has three tiers with
combined biometric modalities that are required to set a human node (see the
`Humanode Biometric Modalities Score' section for more information).

Liveness detection uses different physiological properties to differentiate
between real and fake characters. An AI computer system can determine that
it is interfacing with a physically present human being and not an inanimate
spoof artifact.

A non-living object that exhibits human traits is called an `artifact'. The
goal of the artifact is to fool biometric sensors into believing that they are 4
interacting with a real human being instead of an artificial copycat. When an
artifact tries to bypass a biometric sensor, it is called a `spoof'.
Artifacts include photos, videos, masks, deepfakes, and many other
sophisticated methods of fooling the AI. Another method of trying to
bypass sensors is by trying to insert already captured data into a system
directly without camera interaction. The latter is referred to as `bypass'.

In the biometric authentication process, liveness data should only be
valid for a set period of time (up to several minutes) and then is deleted.
As this data is not stored, it cannot be used to spoof liveness detection
with corresponding artifacts to try and bypass the system.

The security of liveness detection is very dependent on the amount
of data it is able to detect. That is why low-resolution cameras might
never be totally secure. For example, if we take a low-res camera and
put a 4k monitor in front of it, then weak liveness detection methods
such as turning your head, blinking, smiling, speaking random words, etc.
can be easily emulated to fool the system.

In 2017, the International Organization of Standardization (ISO) published
the ISO/IEC 30107-3:2017 standard for presentation attacks and went over
ways to stop artifacts such as high-resolution photos, commercially
available lifelike dolls, silicone 3D masks, etc. from spoofing real identities.
Since then, sanctioned PAD (presentation attack detection) tests for biometric
authentication solutions have been created so that any new solutions meet the
specified requirements before hitting the market. The most famous of them all
is the iBeta PAD Test. It is a strict and thorough evaluation of biometric
processing solutions to see whether they can withstand the most intense
presentation attacks. Four years have passed since then and this standard is
condemned as outdated by many specialists in the field, and iBeta PAD tests
have gradually become easier to pass with modern sophisticated spoofing methods.

FaceTec, one of the leading companies in liveness detection, divides attacks
into five categories that go way beyond those stated in the 30107-3:2017
standard and represent real-world threats much more precisely.

Depending on the artifact type, there are three levels of PAD attacks:

\begin{description}
  \item Level 1—Hi-Res digital photos, HD videos, and paper masks

  \item Level 2—Commercially available lifelike dolls, latex and silicone 3D masks

  \item Level 3—Ultra-realistic artifacts like 3D masks and wax heads
\end{description}

Furthermore, depending on the bypass type, FaceTec researchers identify Level
4 \& 5 biometric template tampering, and virtual camera and video injection attacks:

\begin{description}
  \item Level 4—Decrypt \& edit the contents of a 3D template to contain
        synthetic data not collected from the session, having the server
        process and respond with `Liveness Success'.

  \item Level 5—Take over the camera feed and inject previously captured
        video frames or a deepfake puppet that results in the AI responding with
        `Liveness Success'.
\end{description}

\begin{figure}[H]
  \centering
  \label{fig16}
  \includegraphics[width=0.75\linewidth, keepaspectratio]{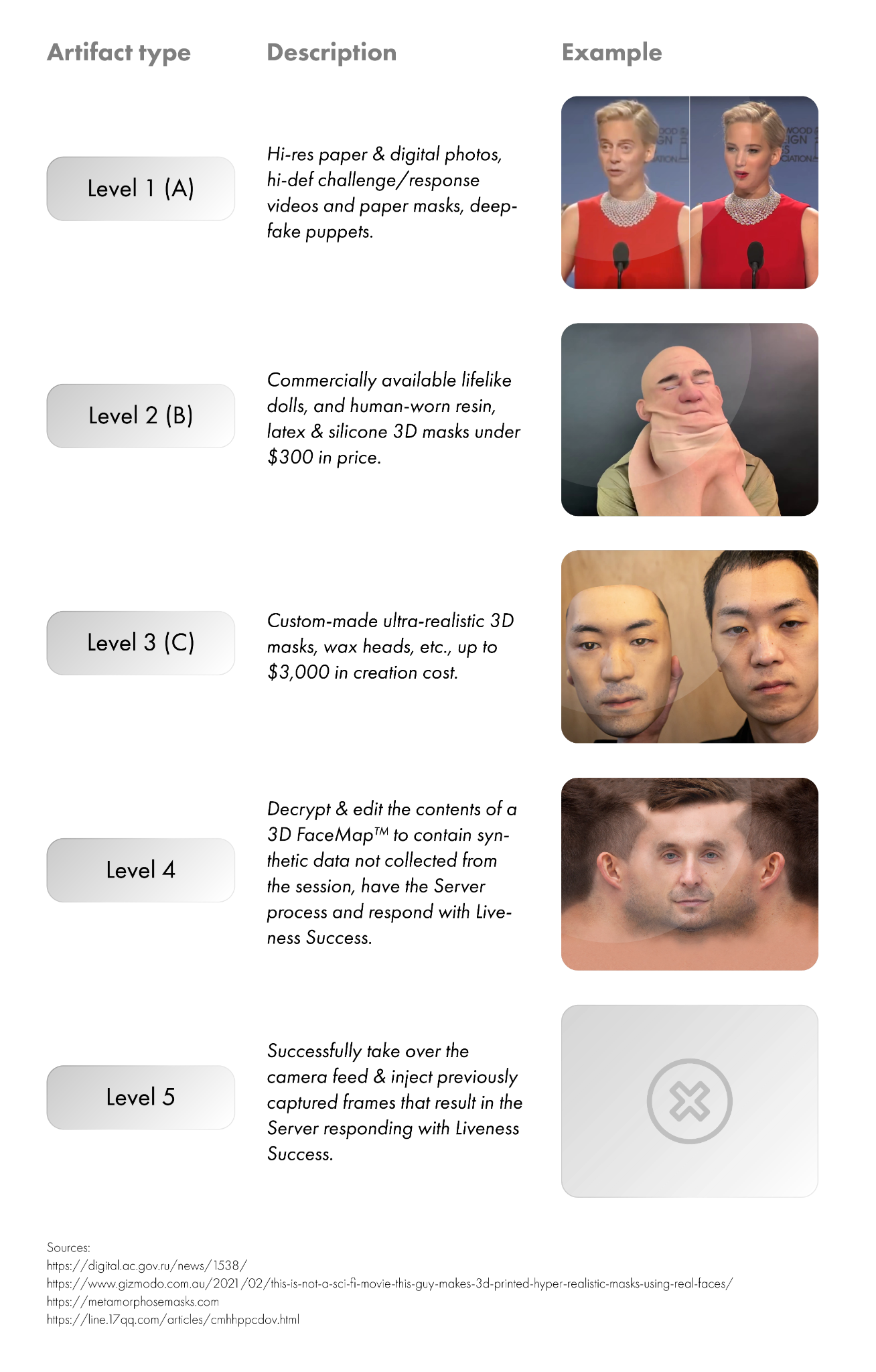}
  \caption{5 levels of liveness.}
\end{figure}

Almost all liveness detection methods as well as those described above in the
Humanode approach to user identification are \emph{software-based} and available
on any modern smartphone. In \emph{hardware-based methods}, an additional device
is installed on the sensor to detect the properties of a living person: fingerprint
sweat, blood pressure, or specific reflection properties of the eye.

With liveness detection, the chances of successful spoofing become low
enough to make the cost of an attack higher by an order of magnitude in
comparison to the potential transaction fees collected by an artificially
created human node minus the costs to run the node.

The Humanode network implements 3D facial liveness detection from the testnet.

\subsubsection{Replay attack}

A replay attack is an attack on the communication channel between the
sensors and the feature extractor module. In this attack, an impostor
can steal biometric data and later can submit older recorded data to
bypass the feature extraction module \cite{Jain2008}.

Traditional solutions to prevent this kind of attack are as follows.

\begin{itemize}
  \item Steganography is the way by which biometric characteristics can be
        securely communicated without giving any clue to the intruders.
        It is mainly used for covert communication and therefore biometric
        data can be transmitted to different modules of the biometric system
        within an unsuspected host image.

  \item Watermarking is a similar technique where an identifying pattern
        is embedded in a signal to avoid forging. It is a way to combat
        replay attacks, but only if that data has been seen before or the
        watermark cannot be removed.

  \item A challenge-response system, in which a task or a question
        is given to the person as a challenge and the person responds
        to the challenge voluntarily or involuntarily \cite{Kamaldeep2011}.
\end{itemize}

\subsubsection{Attack on the channel between the database and the matcher}

The attacker intrudes on the channel to modify the existing data or to
replay the old one. Traditionally, this attack can be prevented by
such solutions as challenge-response systems, watermarking, and steganographic
techniques as a replay attack \cite{Bolle2002}.

\subsubsection{Attack on the database}

The attacker can intervene in the database where the templates are
stored to compromise the biometric characteristics of a user, replace,
modify, or delete the existing templates.

There are two common template protection schemes to counter this attack:

\begin{itemize}
  \item \emph{Cancelable biometrics}, in which the intruder cannot get
        access to the original biometric pattern from the database because
        instead of the original data, a distorted version is stored

  \item \emph{Cryptobiometrics}, where all data are encrypted before
        sending in the database while the original template is deleted,
        therefore, it is quite difficult for the attacker to steal the original
        template, as it exists only for a few seconds on the user's device.
\end{itemize}

The Humanode network uses the second type of template protection scheme.

\subsubsection{Overriding the final decision}

As the software application may have bugs, an intruder can override the
actual decision made by the matcher.

Humanode ensures that nobody knows the actual decision result of
matching but the protocol before this decision is executed. This
kind of attack can be prevented using soft biometrics as well \cite{Kamaldeep2011}.

\subsubsection{Overriding the feature extractor}

This attack relates to overriding the feature extractor to produce
predetermined feature sets, as the feature extractor is substituted and
controlled remotely to intercept the biometric system.

In the Humanode system, feature extraction takes place on the client's device.
The human node encrypts the embedded feature vector using the public key
and gets the encrypted feature vector. Further, it provides ZKP proof
that the feature vector is extracted through the system's feature
extraction process only, as a result the attacker is unable to override it.

\subsubsection{Override matcher}

Overriding the matcher to output high scores compromises system security.
In this way, the intruder can control the matching score and generate a
high matching score to confirm authentication to the impostor.

In Humanode, the matching score is computed over an encrypted feature vector.
Moreover, the matcher is required to provide proof of correctness for the
matched score. As a result, the attacker cannot override matchers to
generate a high matching score for a target feature vector.

\subsubsection{Synthesized feature vector}

The route from the feature extractor to the matcher is intercepted to
steal the feature vector of the authorized user. Using the legitimate
feature vector, the attacker then iteratively changes the false data,
retaining only those changes that improve the score until an acceptable
match score is generated and the biometric system accepts the false data.
The legitimate feature sets are replayed later with synthetic feature sets
to bypass the matcher \cite{Bolle2002, Jain2008, Kamaldeep2011}.

In the Humanode system, there is a private channel between the feature
extractor and the matcher while the feature vector is always kept in encrypted
form and is never available in a plain form to the attacker. Therefore these
kinds of attacks are not possible.

\subsubsection{Reconstruction attack}

Recently, Mai G. et al. \cite{Guangcan2017} proposed a neighborly de-convolutional
neural network (NbNet) to reconstruct face images from their deep templates.
In a distributed P2P network, a node can have access to a biometric template
database, and it can use NbNet to reconstruct corresponding 2D or 3D masks
with a very high success probability for verification.

A robust liveness detection prevents the use of a reconstructed 2D or 3D mask,
but it does not protect the privacy of the corresponding user. For protecting
privacy, there are several solutions based on user-specific randomness
in deep networks and user-specific subject keys. Along with using robust
liveness detection, Humanode stores all biometric templates in the encrypted
form and these are never available in the plain form to the attacker.

\begin{table}[H]
  \centering
  \caption{Attacks on biometric systems and their possible solutions}
  \label{table6}
  \resizebox{\textwidth}{!}{
    \begin{tabular}{|p{5cm}|p{6cm}|}
      \hline
      Attacks                    & Humanode’s approach                                                 \\
      \hline
      Attack on the sensor
                                 & Liveness detection, multimodal biometric systems, soft biometrics   \\
      \hline
      Replay attack              & Steganography, watermarking  techniques, challenge-response systems \\
      \hline
      Attack on the channel between the matcher and database
                                 & Steganography, watermarking techniques, challenge-response systems  \\
      \hline
      Attack on the database     &
      ElGamal Cryptosystem and BFV lattice-based encryption for biometric data                         \\
      \hline
      Override final decision    & Soft biometrics                                                     \\
      \hline
      Override feature extractor & Impossible due to ZKP                                               \\
      \hline
      Override matcher           & Encrypted Matching process                                          \\
      \hline
      Synthesized feature vector &
      A private channel between the feature extractor and the matcher                                  \\
      \hline
      Reconstruction Attack      & Liveness detection, encrypted biometric templates                   \\
      \hline
    \end{tabular}}
\end{table}

\subsection{Neurosignatures}

With the evolution of neural implants, it became possible to convert the
neuroactivity of the brain into electronic signals that can be comprehended by
modern computers. Since the 1960s, the neurotech field has moved from simple
electroencephalography (EEG) recordings to real brain-computer communication
and the creation of sophisticated BCI-controlled applications. Since the late
2010s, large companies have begun to actively pursue BCI development, rapidly
approaching adoption. Companies, like BrainGate and Neuralink, \cite{braingate,neuralink}
have manufactured working prototypes of invasive and non-invasive
BCIs that build a digital link between brains and computers. In 2014,
Brainlab \cite{brainlab} developed a prototype that allows a Google Glass user to interface
with and give commands to the device using evoked brain responses rather than
swipes or voice commands. In 2015, Afergan et al. developed an fNIRS-based BCI
using OST-HMD called Phylter, a control system connected to Google Glass
that helped prevent the user from getting flooded by notifications. In 2017,
Facebook announced the BCI program \cite{Facebook2017}, outlining its goal to build a non-invasive,
wearable device that lets people type by simply imagining themselves talking.
In March 2020, the company published the results of a study \cite{Makin2020} that set a new
benchmark for decoding speech directly from brain activity. Even with the
immeasurable complexity of neurons and ridiculous entanglements of somas,
axons, and dendrites, the above-mentioned projects were able to create
devices that not only stimulate and capture the output but also distinguish
patterns of signals from one another.

A person will be able to use his own mental state, conscious state, or
simply signals from the motor cortex to initiate node deployment and verify
transactions without compromising the data itself.

Compared to any other biometric solution including direct DNA screening
and other biochemical solutions, neurosignature biometrics can be considered
to be the most secure way of biometric processing, as it is impossible to forge a
copycat or to emulate the prover and try to bypass the system.

\begin{figure}[H]
  \centering
  \label{fig17}
  \includegraphics[width=0.75\linewidth, keepaspectratio]{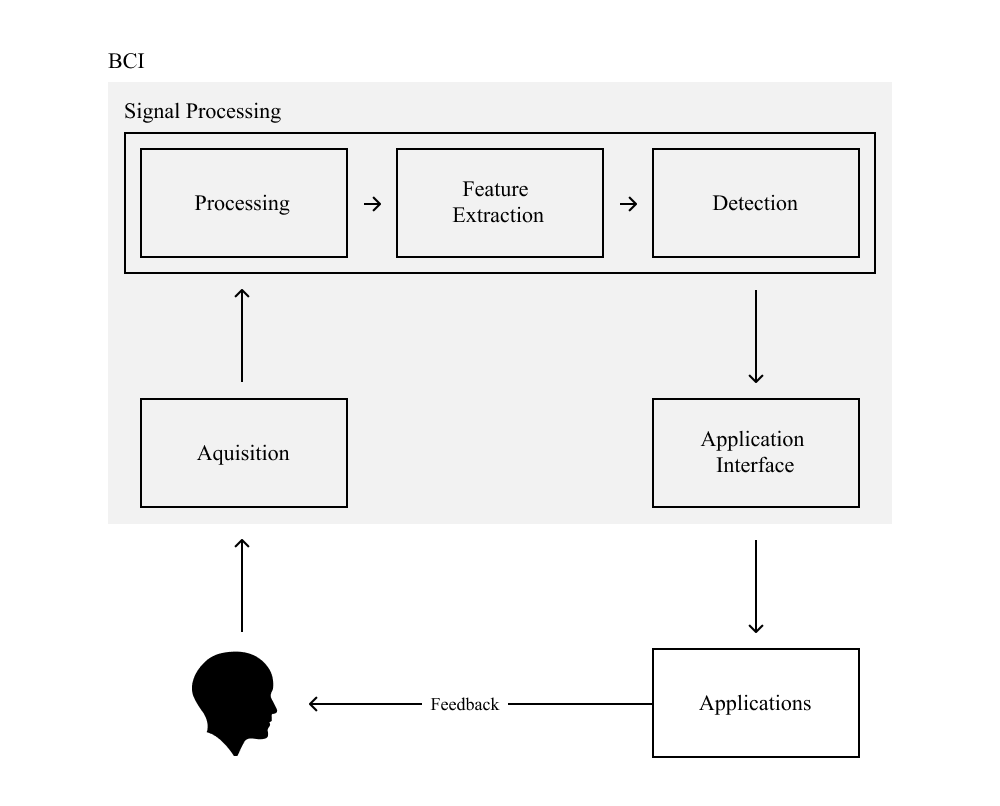}
  \caption{Block diagram of a BCI system.}
\end{figure}

\begin{table}[H]
  \centering
  \caption{Summary of signal acquisition method \cite{Mudgal2020}.}
  \label{table7}
  \resizebox{\textwidth}{!}{
    \begin{tabular}{|c|c|c|c|c|c|}
      \hline
      Types                         & Example        & Signal Type & Portability  & Spatial Resolution & Temporal Resolution \\
      \hline
      Invasive                      & Intra-cortical & Electrical  & Portable     & Very high          & High                \\
      \hline
      Semi-invasive                 & ECoG           & Electrical  & Portable     & High               & High                \\
      \hline
      \multirow{4}{*}{Non-invasive} & EEG            & Electrical  & Portable     & Low                & Mediate             \\
                                    & MEG            & Magnetic    & Non-portable & Mediate            & Mediate             \\
                                    & fMRI           & Metabolic   & Non-portable & High               & Low                 \\
                                    & fNIRS          & Metabolic   & Portable     & Mediate            & Low                 \\
      \hline
    \end{tabular}}
\end{table}

While BCI hardware enables the retrieval of brain signals, BCI software is
required to analyze these signals, produce output, and provide feedback.

Moreover, neurotech continues to evolve—hybrid BCIs (hBCIs),
which are the combinations of BCIs with a wide range of assistive devices, p
rove it \cite{Pfurtscheller2010, Muller2013, Choi2017, Yang2020}. These hBCI systems are categorized according to the
type of signal combined as well as the combination technique
(simultaneous/sequential). Electroencephalography (EEG), due to its easy
use and fast temporal resolution, is most widely utilized in combination
with other brain/non-brain signal acquisition modalities, such as functional
near-infrared spectroscopy (fNIRS), electromyography (EMG),
electrooculography (EOG), and eye-tracking technology
\cite{Hong2017}. In general,
the essential goal of combining signals is to increase detection accuracy,
enhance system speed, improve the user experience, and overcome the disadvantages
of BCI systems \cite{Sadeghi2018}. With hBCIs, Humanode can achieve
unprecedented multi-modality based on internal biometric processing protocols.

There are many different ways to collect data on brain activity, but more importantly,
there have been many software layers already created by different organizations and
communities such as OpenBCI, BCI2000, NFBLab, PsychoPy, rtsBCI, OpenVibe, OpenEEG,
BF++, etc.

These types of software can be divided into three different groups:

\begin{enumerate}
  \item Software that provides a stack of protocols that try to precisely read,
        analyze, and store brain activity data through different types of signals
        (EEGs, fMRIs, invasive implants, etc.)

  \item Software that converts brain activity data into commands for different
        computer languages and systems, and

  \item Supplementary software that converts received brain
        activity data into different types of variables for research
        and development purposes.
\end{enumerate}

\begin{figure}[H]
  \centering
  \label{fig18}
  \includegraphics[width=0.75\linewidth, keepaspectratio]{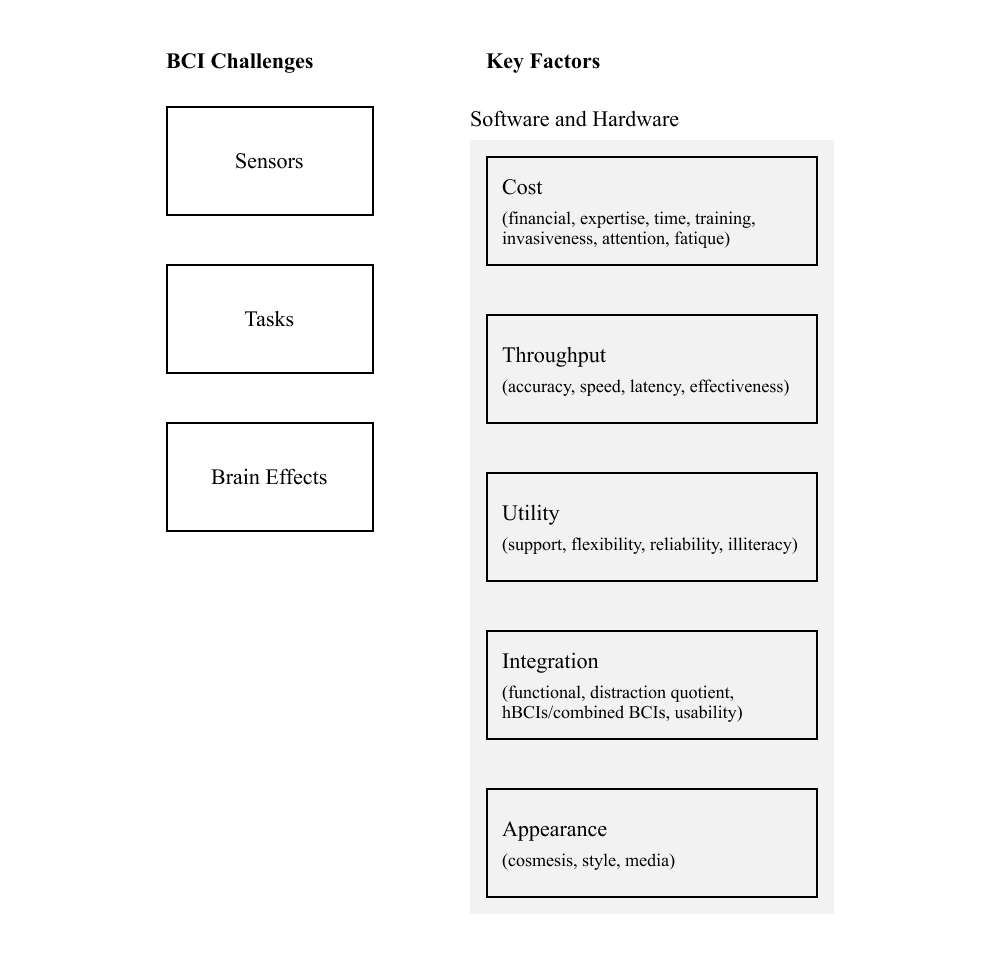}
  \caption{Factors that influence adoption. }
\end{figure}

Researchers are making great strides toward resolving all of the above-mentioned
challenges. The majority of investigators believe in BCI mass adoption in the
following years. Recent research examines the possibility of using BCI in
everyday-life settings in different contexts \cite{Blum2012, Hu2019, Park2020, Friedman2020, Benitez2020}.
There is a relevant body of work addressing not only technology improvements
\cite{Liberati2015} but also the fact that BCI design and development
should become more user-friendly to achieve successful mainstream applications
\cite{Kubler2014, Nijboer2015}.

\subsection{Humanode's approach to identity attack prevention}

The amount of research exploring the use of distributed ledger technology to
launch new types of identity management systems has lately increased
\cite{Baars2016, Jacobovitz2016, Tobin2016, Dunphy2018}, along with studies combining
these systems with biometrics \cite{hammudoglu2017, Garcia2018, Othman2018}.

\begin{figure}[H]
  \centering
  \label{fig19}
  \includegraphics[width=0.75\linewidth, keepaspectratio]{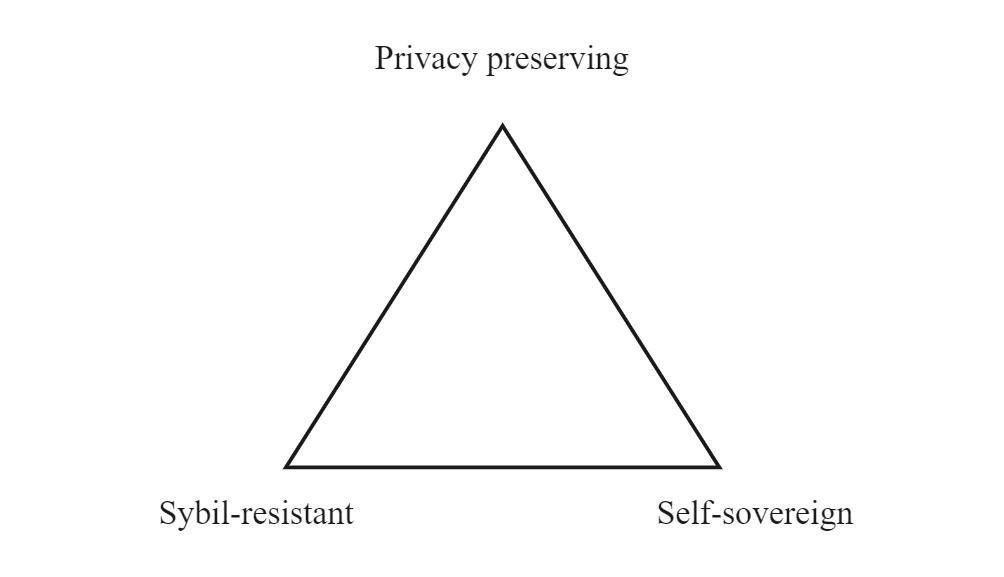}
  \caption{Decentralized Identity Trilemma \cite{maciek2018}.}
\end{figure}

Alongside maintaining \emph{self-sovereignty} (anybody can create and control an
identity without the involvement of a centralized third party) and being
\emph{privacy-preserving} (anybody can acquire and utilize an ID without revealing PII),
the system also needs to achieve \emph{Sybil resistance}, as the majority of large-scale
peer-to-peer networks are still vulnerable to Sybil attacks. These occur
where a reputation system is subverted by a considerable number of forging
IDs in the network \cite{Douceur2002, RincyMedayil2015, BHISE2016, Siddarth2020}.

None of the existing solutions are privacy-preserving, self-sovereign, and
Sybil-resistant at the same time \cite{maciek2018}. We at Humanode propose the
following solutions to break the trilemma.

\subsubsection{Self-sovereignty}

The Humanode protocol applies principles of self-sovereign identity,
requiring that users be the rulers of their own ID \cite{Allen2016}.
In Humanode, there is no centralized third party to control one’s ID,
thus ID holders can create and fully control their identities.

\subsubsection{Privacy-preserving}

In order to meet the security requirements of protecting highly private
biometric information on a truly global decentralized system that is run
on nodes by everyday people, simply encoding the information using
cryptography (no matter how high the encryption) is not enough.
We also need to consider the integrity of the information, preventing
malicious actors from accessing the information and the network as a whole,
preventing Sybil attacks, deepfakes, and an endless number of various possible
and potential attacks. This is where the concept of cryptobiometrics comes into play.

Obviously, in order to safeguard the information while allowing the necessary
information (such as if this is a registered user or not, or what account is
he or she tied to), cryptobiometrics is based on a combination of various
technologies and exists at the intersection of the disciplines of mathematics,
information security, cybersecurity, Sybil resistance, biometric technology,
liveness detection, zero-knowledge–proof (ZKP) technologies, encryption, and
blockchain technology.

\subsubsection{Sybil resistance}

A Sybil-proof system was best conceptualized by Vitalik Buterin as a
``unique identity system'' for generating tokens that prove that an identity
is not part of a Sybil attack \cite{Buterin2014, Buterin2019}. In recent years,
attempts in the field were made by blockchain-based initiatives like
HumanityDAO, POAP, BrightID, Idena Network, Kleros, Duniter, etc.
Nevertheless, there are still no relevant Sybil-resistant identity
mechanisms. In other words, in today’s digital space a possibility
remains for users to create multiple accounts in one system using distinct
pseudonyms to vote several times or receive multiple rewards, etc.

\begin{table}[H]
  \centering
  \caption{Comparison of Sybil attack types.}
  \label{table8}
  \resizebox{\textwidth}{!}{
    \begin{tabular}{|c|p{6cm}|p{4.5cm}|}
      \hline
      Type of Attack            & Description                                             & Defense Method \\
      \hline
      Routing                   & These attacks include distortion of routing protocols:
      single multiple paths through Sybil nodes, or geographic routing,
      in which sensor nodes send data to a base station.
                                & Graph-based detection methods                                            \\
      \hline
      Distributed Storage       & An attacker stores data about false IDs and manipulates
      users to store data in multiple Sybil IDs of a network node.
                                & Machine-learning techniques                                              \\
      \hline
      Data Aggregation          & An attacker uses multiple IDs and modifies aggregation
      readings in the sensor network as a strategy to save energy.
                                & Machine-learning techniques                                              \\
      \hline
      Voting/Reputation Systems & An attacker manipulates systems that use
      voting to accept false solutions and affects the ranking mechanism
      in reputation systems.
                                & Graph-based detection methods                                            \\
      \hline
      Resource Allocation       & These attacks are common in networks where
      resources are assigned depending on the number of nodes.
      Malicious nodes can deny legitimate ones from accessing network resources.
                                & Prevention schemes and graph-based detection methods                     \\
      \hline
      Misbehavior Detection     & An attacker creates multiple Sybil nodes to
      spread false alarms to impact system performance and compromise detection accuracy.
                                & Graph-based detection and manual verification                            \\
      \hline
    \end{tabular}}
\end{table}

\begin{figure}[H]
  \centering
  \label{fig20}
  \includegraphics[width=0.75\linewidth, keepaspectratio]{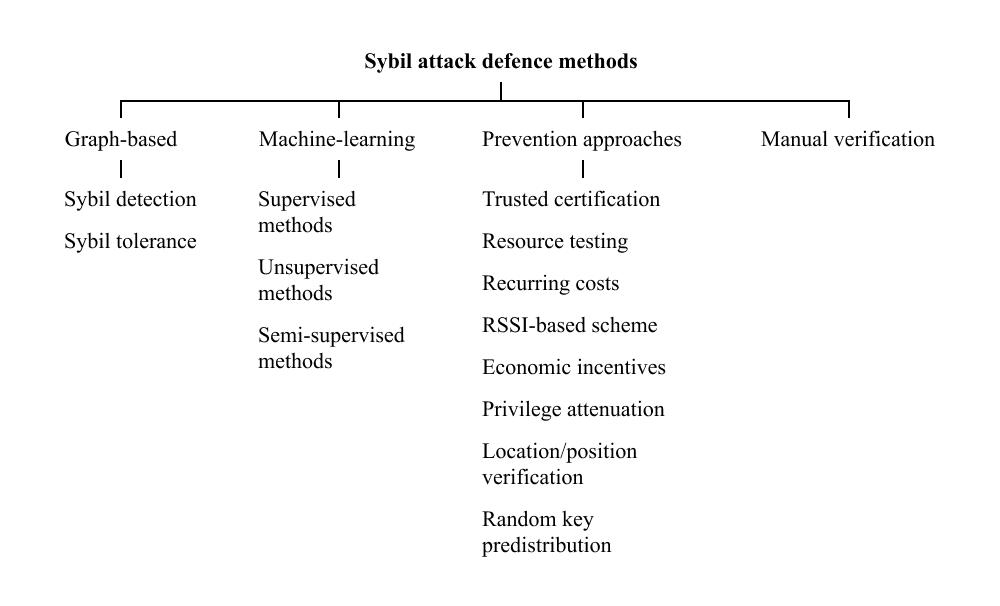}
  \caption{Main Sybil attack defense methods.}
\end{figure}

\begin{itemize}
  \item \emph{Graph-based methods}
\end{itemize}

Graph-based methods rely on a social network's information to represent
dependencies between objects. These schemes fall into two categories:

\begin{enumerate}
  \item Sybil detection techniques based on the concept of graph random walk
        and mix time

  \item Sybil tolerance techniques, which limit the effects of Sybil
        attack edges \cite{Qurishi2017, Alharbi2019}.
\end{enumerate}

\begin{itemize}
  \item \emph{Machine-learning methods}
\end{itemize}

These methods fall into the following categories:

\begin{enumerate}
  \item Supervised, which use regression models, support vector machine (SVM)
        \cite{Pengwenlong2017}, and decision tree models

  \item Un-supervised, which use fuzzy logic, Markov models
        \cite{Zhang2015}, and clustering methods

  \item Semi-supervised, which use sets of data to improve the quality of learning.
\end{enumerate}

\begin{itemize}
  \item \emph{Manual verification methods}
\end{itemize}

This scheme relies on users to increase security through user verification, e.g.,
this may include asking users to report malicious content in the network.

\begin{itemize}
  \item \emph{Prevention methods}
\end{itemize}

Prevention schemes refer to such traditional approaches as using trusted
authorities or resource testing. They may also include the use of
crypto puzzles (CAPTCHA) for users to access systems and verifying
their ID by sending a verification SMS message to the user’s phone.

Humanode uses various techniques for preventing Sybil attacks:

\begin{table}[H]
  \centering
  \caption{Main techniques for preventing Sybil attacks.}
  \label{table9}
  \resizebox{\textwidth}{!}{
    \begin{tabular}{|c|c|p{7.5cm}|}
      \hline
      Technique           & Application Domain                                                     & Description \\
      \hline
      Cryptobiometrics    & General
                          & A combination of cryptographically secure matching and liveness
      detection mechanisms to verify the uniqueness and existence of real human beings.                          \\
      \hline
      Resource Testing    & General
                          & This verification method aims to determine if the identity has as many
      resources as the single physical device it is associated with.                                             \\
      \hline
      Recurring Costs     & General
                          & This technique is a form of resource testing where resource
      tests are performed at regular time intervals to impose a certain ``cost''
      on the attacker that is charged for every identity that she controls or
      introduces into the system \cite{Muliadi2003, Awerbuch2004, Maniatis2003}. However, these researchers
      have used a computational power in their resource test that may not be
      sufficient to control the attack, since the attacker only incurs a one-time
      cost that can be recovered via the execution of the attack itself
      \cite{Levine2005}.                                                                                         \\
      \hline
      Economic Incentives & General
                          & This technique is based on a scheme where economic incentives are
      used to reward the adversaries if the identities that are controlled
      by it are revealed \cite{Margolin2007, SHAREH2019, Shareh2020}.
      The main disadvantage is that it may encourage attackers economically.                                     \\
      \hline
    \end{tabular}}
\end{table}

From the very start, Humanode uses the above-mentioned prevention
methods to successfully counter Sybil attacks. Also, imposing economic
costs as barriers to becoming a human node are used in the system to
make attacks more expensive and less feasible.

In order to create a Sybil-resistant system for human identification,
Humanode ensures that every identity is:

\begin{itemize}
  \item \emph{Unique} (two individuals should not have the same ID)

  \item \emph{Singular} (one individual should not be able to obtain more
        than one ID; \cite{Wang2020})

  \item \emph{Existing} (the person behind the ID is alive and well)
\end{itemize}

To validate users' identities and to create a Sybil-proof system,
Humanode introduces a verification mechanism when the identity is
derived from one or more unique features of the human body—with the
implementation of premiere biometric solutions such as:

\begin{itemize}
  \item Multimodal biometric processing with liveness detection and periodic
        verification of identity

  \item Biochemical biometrics—direct DNA screening and neurosignature
        biometrics through BCI
\end{itemize}

\begin{center}
  ***
\end{center}

Thus, in a nutshell, Humanode’s identity attack prevention scheme solves
Maciek’s ‘Decentralized Identity Trilemma,’ as the system applies
self-sovereignty, privacy-preservation, and Sybil-resistance principles as
illustrated below.

\begin{figure}[H]
  \centering
  \label{fig21}
  \includegraphics[width=0.75\linewidth, keepaspectratio]{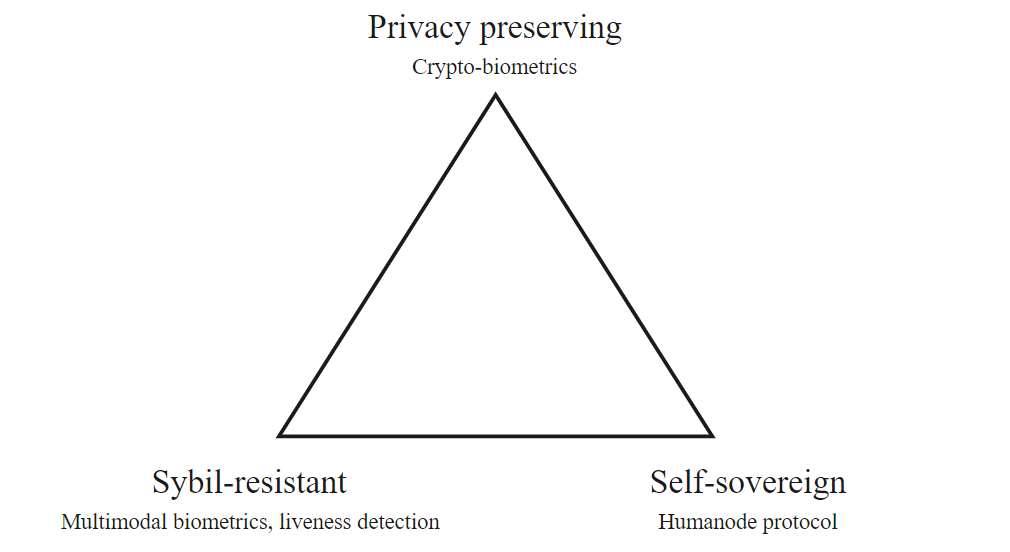}
  \caption{Humanode’s approach to identity attack prevention}
\end{figure}

\section{Transaction fee economics of the Humanode network}

This part of the paper describes important considerations when
selecting an optimal fee strategy for a public cryptonetwork.
It is proposed that the Humanode network will have a cost-based
fee policy that differentiates it from the existing public
permissionless systems based on the internal gas market.

\subsection{Capital-based consensus mechanisms}

Proof-of-Work (PoW) and Proof-of-Stake (PoS) networks, despite
being public and permissionless, build trust through capital
requirements—running devices or acquired and locked tokens.
Hence, they are always susceptible to a direct attack that
needs only capital or to an attack of the external
network \cite{ford2019} of a greater hashing
power or capitalization.

For this reason, the main principle behind the Humanode
network is equal control of the shared truth among each
person joining the system where one cannot achieve additional
voting power toward the consensus of global truth through
money or authority. One living person can launch only one node.

\subsection{Cost-based fee system}

To overcome the problems that current public permissionless
networks face, we apply a cost-based approach to set
transaction fees. This enormously decreases the influence
of the market on the base transaction cost making it more stable over time.

Storage and computing are commodities. Web service providers
with data centers across the globe quote their prices openly.
The amount of Humanode tokens (HMND) that are spent on
renting computing resources from the largest web service
provider in a certain period determines the amount of HMND
tokens the user pays after submitting the transaction.
At the same time, the total computational costs the network
incurred for processing a user's transaction is the
computational transaction fee of the user.

The protocol gets the computational cost in USD quoted
by major cloud computing and storage providers through an
oracle-checked API. The largest price for renting hardware
will be set as the base computational cost of the node.

Instead of having an internal gas token with its own market,
in a cost-based fee system the currency most used by cloud
service providers becomes the internal non-tradable fee token.
For US dollars we use gusd as an internal gas token.
It is used only for determining the protocol’s fees at an
exact moment in time. Actual fees are paid in HMND, and the amount
is defined by the exchange rate provided by the oracles
including decentralized exchange data.

In order to define a user's fee, we multiply the validator's
cost of computation by the number of validators involved in
processing the data.

Now after agreeing on the transaction, human nodes have to
store the updated ledger on the rented or self-launched server.
Storage costs are time-based. Knowing the market price per GB
per month quoted by the leading web service provider,
we set the base cost for a set time period.

We know the storage costs for a given month.
However, the price of the storage resources does
not stay the same over a longer time frame. Hence,
we have to define a formula that determines the cost
of permanent storage over time. This formula is based on
the data sources, methodology and, eventually, rate of cost
decline and is agreed upon in the Vortex.

Over the past 50 years, the cost of commercially available
storage has been decreasing at an annual rate of
30.57\% \cite{Williams2019}. Extrapolating from this data,
the costs of perpetual data storage can be presented as the
infinite sum of the declining storage costs over time:

$$P store = \sum_{i=0}^{\infty} (Data_{size} \ast P_{GBH}[i])$$

where

$ P Store = $ The perpetual price of storage

$ P_{GHB}[i] = $ The cost of storing $1$ GB for an hour at time $i$

$Data_{size} = $ The quantity of data to store

Based on these costs, we derive a transaction fee the user
will have to provide to the Humanode protocol for perpetual
storage of the transaction.

Submitting the transaction, the user pays the cost of perpetual
storage multiplied by the number of nodes storing the data
at the moment. After receiving the fees, the protocol
distributes them equally among the human nodes.
This mechanism ensures that the fee the protocol gets is higher
than the actual cost for storing data, which includes its
future storage. This creates an incentive to continue
operating a Humanode server, ensuring the network’s long-term stability.

Not all of the transactions are equal. \emph{The size of the
  transaction changes with the amount of data and computation
  complexity}. The core functions of the Humanode protocol are
to encrypt, process, and store biometric data, send
transactions with on-chain assets, execute smart contracts,
and connect to other ledgers and databases.

\section{Vortex}

The concept of a ``Decentralized Autonomous Organization''
(DAO) was proposed by Daniel Larimer (\cite{Larimer2013}) and implemented
in Bitshares \cite{Bitshares} in 2014. In 2014, Vitalik Buterin, the founder
of Ethereum proposed that after a DAO was launched, it
might be organized to run without human managerial interactivity,
provided the smart contracts were supported by a
Turing-complete platform \cite{Buterin2014-2}.
Thereby, DAO clearly designates something broader than the
typical definition of “organization”—a social group that
brings people together and works toward a common purpose.
Vitalik thus defines a DAO as ``a decentralized autonomous community''
in which all members have a share in the decision
making \cite{Buterin2013}, that is ``an entity that lives
on the internet and exists autonomously, but also heavily
relies on hiring individuals to perform certain tasks
that the automaton itself cannot do'' \cite{Buterin2014-2}.

Governance in the Humanode network will be decentralized
from genesis and is known as Vortex—the Humanode DAO,
to propose and decide on changes to the Humanode network.
Vortex consists of human nodes, Delegators, and Governors.

\begin{enumerate}
  \item \emph{Human node}—a user who has gone through
        proper biometric processing and receives network transaction
        fees but does not participate in governance.
  \item \emph{Governor}—a human node who participates
        in voting procedures according to governing requirements.
        If governing requirements are not met, the protocol
        converts him back to a non-governing human node automatically.
  \item \emph{Delegator}—a Governor who decides to delegate
        his voting power to another Governor.
\end{enumerate}

The Governors will have different rights according to their tiers.
Tiers are based on Proof-of-Time (PoT) and Proof-of-Devotion (PoD),
meaning that devotion in the system is valued more than the riches one has.
Tiers do not give any additional voting powers to their holders;
instead, they are given the ability to make and promote proposals
on crucial matters. If a human node wants to become a Governor,
they must have their proposal accepted by the Vortex DAO.
All proposals are submitted to the proposal pool anonymously.

\begin{table}[H]
  \centering
  \caption{Governor requirements and tiers. }
  \label{table10}
  \begin{tabular}{|p{3cm}|c|c|c|c|}
    \hline
    Req Tier                                      & Citizen      & Senator      & Legate   & Consul   \\
    \hline
    Need to govern, years                         & 0            & 1            & 2        & 4        \\
    \hline
    Participate in Formation                      & Not required & Not required & Required & Required \\
    \hline
    Run a node                                    & Required     & Required     & Required & Required \\
    \hline
    Have one of your proposals approved by Vortex & Required     & Required     &
    Required                                      & Required                                          \\
    \hline
  \end{tabular}
\end{table}

The combination of PoT and PoD in Humanode
governance means that a Governor progresses through
tiers based on the time their node was considered
governing and the amount of devotion a Governor
channeled into the network. On top of that,
to progress to a Legate or higher, a Governor must
participate in Formation, a proposal-based grant mechanism.
Even if a Governor participates in Formation during the first
days of his node's existence, he will still be required to
govern for another three years to become a Legate.

\begin{table}[H]
  \centering
  \caption{Governor rights. }
  \label{table11}
  \begin{tabular}{|p{3.5cm}|c|c|c|c|}
    \hline
    Rights                                & Citizen & Senator & Legate & Consul \\
    \hline
    Vote on proposals                     & Yes     & Yes     & Yes    & Yes    \\
    \hline
    Participate in Formation              & Yes     & Yes     & Yes    & Yes    \\
    \hline
    Nominate proposals of non-human nodes & Yes     & Yes     & Yes    & Yes    \\
    \hline
    Receive voting delegation             & Yes     & Yes     & Yes    & Yes    \\
    \hline
    Make product proposals                & Yes     & Yes     & Yes    & Yes    \\
    \hline
    Fee distribution proposals            & No      & Yes     & Yes    & Yes    \\
    \hline
    Monetary proposals                    & No      & No      & Yes    & Yes    \\
    \hline
    Protocol-level proposal               & No      & No      & Yes    & Yes    \\
    \hline
    Administrative proposal               & No      & No      & Yes    & Yes    \\
    \hline
    Vortex core proposal                  & No      & No      & No     & Yes    \\
    \hline
    Veto                                  & No      & No      & No     & Yes    \\
    \hline
  \end{tabular}
\end{table}

A quorum is reached if at least 33\% of the Governors
vote on a proposal. If 66\% of the Governors within the
quorum vote to approve a proposal, then Vortex will consider it approved.
This means that 22\% of the Governors will be the necessary
minimum to approve a proposal. Human nodes that do not participate
in governance are not counted in reaching a quorum.

The voting power of each Governor is equal to 1 + the votes of his Delegators.

Any proposal that is pulled out of the proposal pool gets a
week to be voted upon in Vortex.

\begin{figure}[H]
  \centering
  \label{fig22}
  \includegraphics[width=0.75\linewidth, keepaspectratio]{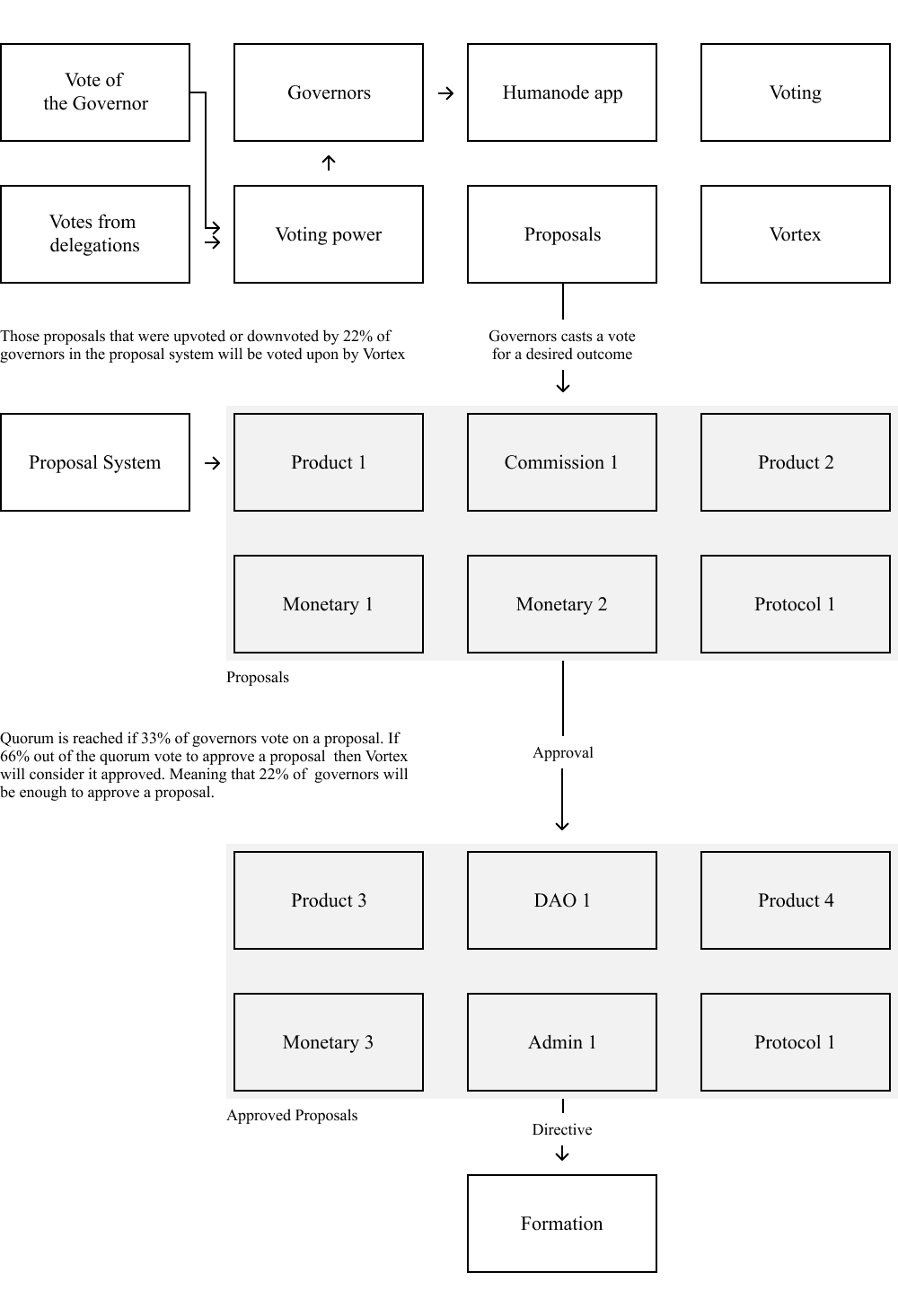}
  \caption{Vortex voting procedures. }
\end{figure}

Becoming a part of Vortex gives access to different governing
tools based on the user’s Governor tier.
Hypothetically, separation of voting powers from proposal
rights that are solely determined by time and participation
should make the whole system reasonably decentralized,
preventing malicious actors from a quick attack on the Humanode network.

\subsection{Tiers and proposal types }

Tiers give various proposal rights to the Governors. The higher the tier the
more crucial a proposal a Governor can make.

In more detail, Humanode Upgrade Proposals (HUPs) are divided into the
following categories that are based on the inception characteristics of Vortex:

\subsubsection*{Tier 1 Citizen}

\paragraph*{Product}

Changes in the products. Max time in proposal pool: 2 weeks.
\begin{itemize}
  \item Logo

  \item Design

  \item Social media presence

  \item Web, mobile, and desktop application for dashboard,
        wallet, biometric verification, and voting

  \item Website, humanode.io domain name

  \item Proposals for new products
\end{itemize}

\subsubsection*{Tier 2 Senator}

\paragraph*{Fee distribution}

Vortex can change fee distributions. Maximum time in the proposal pool: 1 month.
\begin{itemize}
  \item 98\% of the fee is equally given out to every human node;

  \item  2\% of fees flow into the Formation vault to fund the
        network development and execute proposals.
\end{itemize}

\paragraph*{Monetary}

Modifying Humanode's monetary system and its principles via DAO.
Max time in proposal pool: 1 month.

\begin{itemize}
  \item Creation of HMND tokens

  \item Implementation of Fath on the Humanode main net
        \begin{itemize}
          \item Proportional emission distribution

          \item Monetary Supply Balancing mechanisms
        \end{itemize}

  \item Equality of fee distribution among human nodes
\end{itemize}

\subsubsection*{Tier 3 Legate}

\paragraph*{Protocol}

HUPs are the way to create enforceable changes to the Humanode protocol.
Max time in proposal pool: 2 months.

\begin{itemize}
  \item Combination of biometrics through multimodal biometric processing
        in node creation

  \item Substrate blockchain

  \item Consensus mechanism (Aura, Snowball, Grandpa, Nakamoto, Agnostic)

  \item Sybil defense through decentralized cryptobiometrics

  \item Equality between peers in decisions on a global state

  \item Delegation mechanics
\end{itemize}

\paragraph*{Administrative}

Max time in proposal pool: 3 months.
\begin{itemize}
  \item Types of human nodes, their rights, and requirements

  \item Governor tiers: rights, requirements, and obligations

  \item Formation procedures and grants
\end{itemize}

\subsubsection*{Tier 4 Consul}

\paragraph*{Vortex Core}

A DAO begins with a defined scope of proposal types to prevent detrimental actions.
But it is not supposed to stay narrow. The system will eventually allow the
submission of HUPs to do anything possible on the DAO. Simply by submitting proposals,
Vortex can go wherever the imagination takes proposers.
Maximum time in the proposal pool: 6 months.

\begin{itemize}
  \item Proposal system values and protocol

  \item Vortex voting values and protocol

  \item Equal voting power distribution

  \item Decisions on the creation of new types of human nodes

  \item Decisions on the creation of new types of Governors
\end{itemize}

The above-mentioned characteristics will be implemented on top of the Humanode
network at the deployment stage.

\subsection{Veto rights}

If for some reason 66\% percent of Consuls decide that it is necessary to
veto a decision then it will be possible to do so, and the decision will
be considered declined by the Vortex. But they cannot veto \emph{the same decision}
more than two times in a row, meaning that if a proposal is approved
thrice then the veto cannot be implemented. Vetoes are important to
safeguard the system from panic-based attacks and the dilemma where a
minority of professionals might be able to see things clearer than the
whole mass of voters. But liberty, public opinion, and democracy should
prevail in the end, as the Consuls’ veto cannot be implemented more than
two times for a particular decision.

\subsection{Proposal system}

The two main principles behind creating the Humanode proposal system are to
mitigate chokepoints and to keep up the quality of proposals. Governors can
participate in every part of the system while other human nodes can only make
proposals. Non-human nodes cannot propose directly but can be nominated by any
Governor to do so. A human node cannot create more than five proposals at the same time.
All proposals are submitted to the proposal pool anonymously.

\begin{figure}[H]
  \centering
  \label{fig23}
  \includegraphics[width=0.75\linewidth, keepaspectratio]{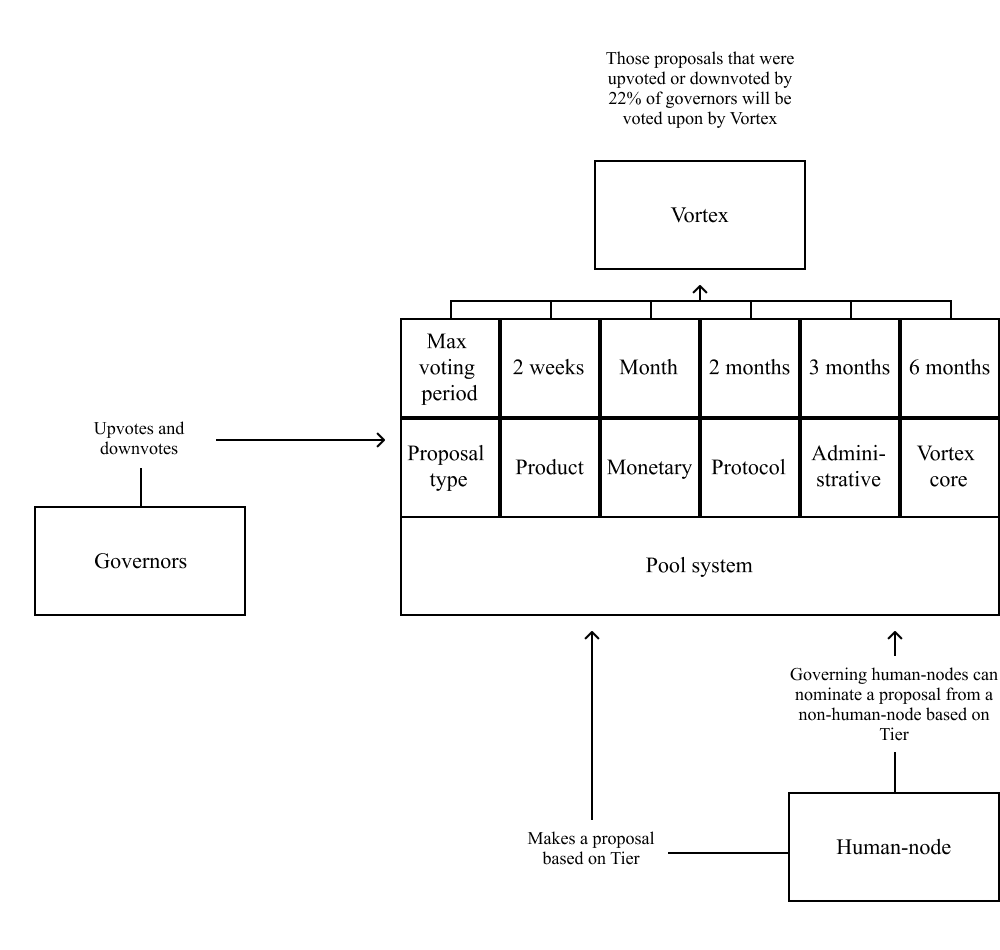}
  \caption{Proposal pool system and voting periods.}
\end{figure}

How it works:
\begin{enumerate}
  \item A human node casts the proposal into the pool system, defining a header,
        the voting period, writing a description, adding docs, etc., but more
        importantly, choosing one of the types of proposals that are available
        depending on the governing tier.

  \item Inside the pool, Governors upvote or downvote different proposals.
        Each Governor can give each proposal an upvote or a downvote. Each pool
        consists of different boards: fresh, trending, popular, new, etc.

  \item Proposals that receive upvotes or downvotes from 22\% of existing
        Governors are immediately conveyed to Vortex for voting.
        Proposals that do not receive enough upvotes or downvotes in the max
        voting period get deleted from the pool and can be proposed again in two weeks’ time.

  \item The voting procedure in Vortex takes up strictly a week for
        each proposal to be voted upon.

  \item If approved the proposal is conveyed to Formation to receive
        funding and assemble a team.

  \item If declined by Vortex, the proposers must wait out a period of
        two weeks to propose again.
\end{enumerate}

\subsection{Formation}

Vortex governs the Humanode by deciding on key parameters through the voting
power of human nodes.

Formation is a part of the Humanode. It is a special grant-based development
system providing grants, investments, service agreements, and projects to build.
It is dedicated to supporting the Humanode network and all related technologies.

Any human node can join Formation to make a grant proposal or apply to
become a part of a team that already develops an approved proposal.
Proposals by non-human nodes can only reach Formation if one of the governing
human nodes decides to nominate them. Such limitations allow us to protect devoted
followers and contributors to the Humanode network.

The process is as follows:

\begin{figure}[H]
  \centering
  \label{fig24}
  \includegraphics[width=0.75\linewidth, keepaspectratio]{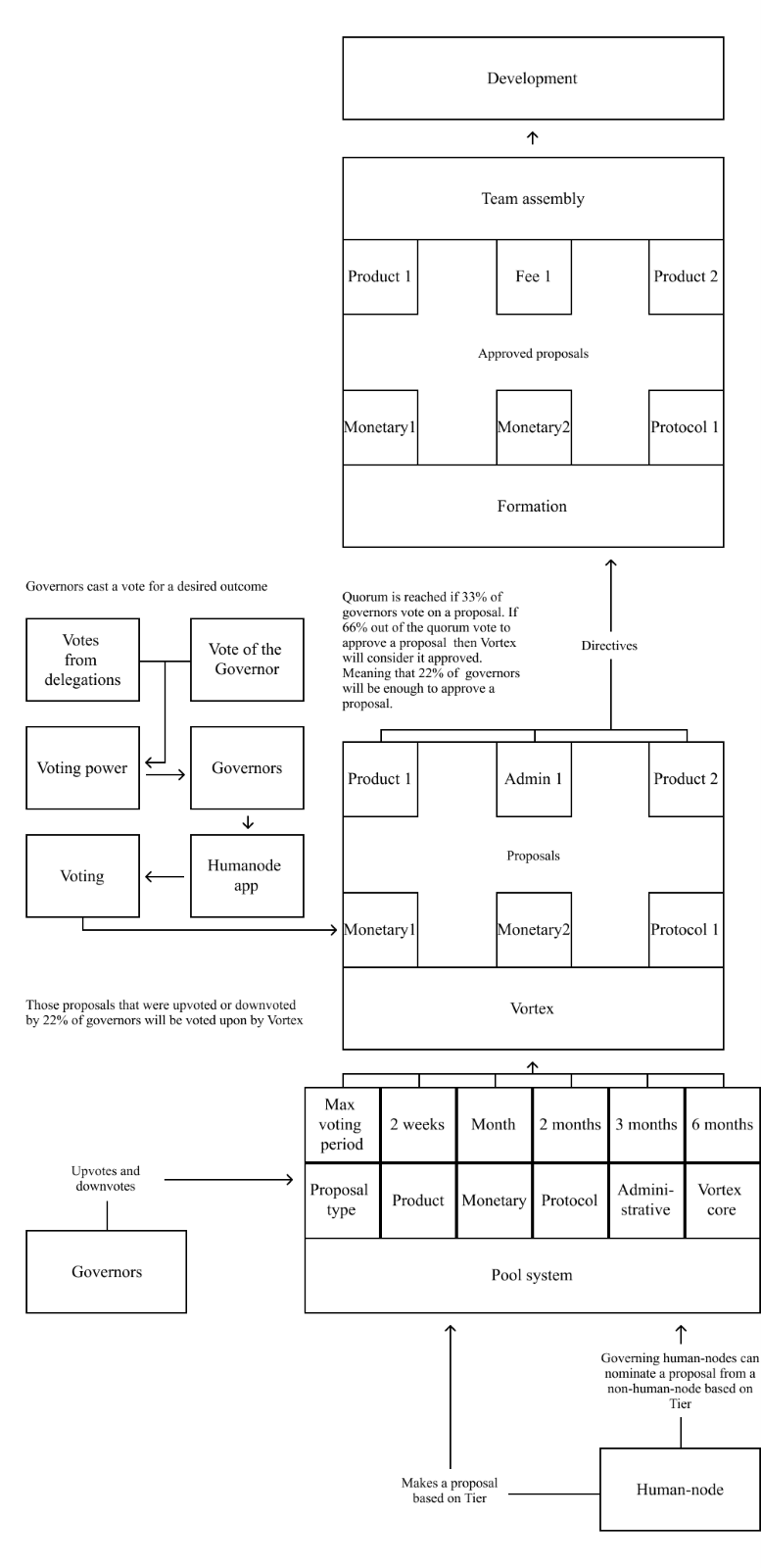}
  \caption{Proposal pool, Vortex, and Formation processes.}
\end{figure}

Human nodes create proposals, allocate funds for their implementation,
and take coordinated action to see the proposals implemented properly.
Governors upvote and downvote them. We assume that 2\% of fees go to
Formation as the network begins to function. Then, the proposers, i.e.,
Vortex, will regularly determine the percentage of the fees going to Formation.

The Humanode network's DAO supports a number of different proposal directions.

Generally, Formation funds:
\begin{itemize}
  \item \emph{Research}: Advancing basic and applied research in
        cryptobiometrics, cryptographic primitives, distributed systems,
        consensus mechanisms, smart-contract layers, biometric modalities,
        liveness detection, encrypted search, and matching operations.

  \item Development and Product: Development turns research into software,
        while Product turns it into user experiences. Formation is primarily
        interested in technologies that expand the Humanode network, its potential,
        capabilities, and security, as well as the ecosystem, from decentralized
        finance and non-fungible tokens to decentralized courts.

  \item Social Good \& Community: Formation supports community members
        to bring awareness to open-source, decentralized networks, and
        biotechnologies, and scale community outreach for the Humanode network.
\end{itemize}

The Formation funds are mainly used to maintain the network.

\subsubsection{Assembling a team}

We understand how crucial it is to find and coordinate people that are
willing to support the proliferation of the Humanode network.
That is why we are developing a special team-assembly procedure in the
Humanode app that will allow those whose proposals were approved by
Vortex to find passionate professionals to assemble their team from
among the members of the international Humanode community.
All the proposer has to do is send a digital offer to any other human
nodes that he thinks are a good fit for his projects. Their proposal must
have the public address of the potential member, and it should state
working objectives and conditions and have a smart contract that locks
some part of the grant for that person in particular.

Proposers—the team (public key/role).
\begin{itemize}
  \item Full team

  \item The team is partially assembled

  \item No one in the team yet
\end{itemize}

If the grant proposal in Formation does not include the team,
its members might be selected from the human nodes who are interested in
the proposed project.

\section{Discussion}

\subsection{Gradual decentralization}

Obviously, the Humanode network relies on the activity of its Governors.
Besides building the technological solutions stated in this paper,
the Humanode core will promote full transparency of governing processes
and transactions, design and deploy decentralized governing processes,
participate heavily in the Humanode community, and make development proposals.
The Proposal Pool System/Vortex–Formation governance stack was designed
by the Humanode core to create a hybrid
Proof-of-Time/Proof-of-Devotion/Proof-of-Human-Existence
safeguarded network. This implementation allows us to lower the
influence of the problems that affect any system that tries to integrate
democratic procedures:

\begin{enumerate}
  \item Voter apathy is a very widespread problem that entangles
        every single voting system. The biggest part of this problem is
        the inability to reach a quorum. The Humanode network demands
        governance participation in proposals and voting from Governors
        and proof of existence from all human nodes. Those Governors
        who do not fulfill monthly governing conditions
        (either they did not make proposals or did not vote on any proposal)
        are automatically converted to non-governing.
        Quorum is reached if 33\% of Governors vote upon a proposal,
        so it means that only voices of those who actively participate
        in governance are calculated to reach a quorum.

  \item Masses are often mistaken. It is common sense that a small,
        dedicated group of professionals with years of experience would be
        able to give a more precise and correct opinion on a particular
        voting matter than a mass of people with different backgrounds
        and education. To balance the democratic approach with
        professional education and experience, Humanode core came up
        with a hybrid Proof-of-Time/Proof-of-Dedication governance system named
        ``Vortex'', in which Governors have different tiers.
        They can be promoted in tiers if certain requirements are met.
        This way the protocol gives more tools and proposal rights to those
        who have more experience and have proven their devotion through Formation.
        The necessity to have your proposal approved before becoming a
        Governor acts as a Proof-of-Devotion step that uplifts the
        quality of Governors and acts as an important layer of
        defense against Sybil attacks.

  \item Inability to directly delegate your vote to any other voter
        in a system creates many different forms of how the voting
        procedures take place. The very systems of how electoral
        delegates are chosen have loopholes that allow political tricks
        such as gerrymandering and filibustering. Governing human nodes
        are designed to be equal in voting power; at the same time, the
        voting mechanisms allow you to delegate your vote to any other
        human node without boundaries. A Governor's voting power equals
        $1 +$ the number of delegations he has.
\end{enumerate}

As we try to balance freedom with safety and quality we were faced
with two options: to either be a centralized company that does
everything on its own until it is working optimally or decentralizing
it and building on a devoted community. The first approach surely
has its advantages in terms of coordination and speed, but it is
authoritarian, and Humanode is not. But if we chose the second option,
the Humanode network would be at a larger risk, as it is going to be
quite small at first. So to solve this dilemma, instead of choosing
one out of the two approaches we came up with a third solution.
We decided that all members of Humanode core will receive a
Consul tier at the deployment phase so that we as founders and
developers would have the ability to lead a more centralized approach
in governance at first. Decentralization is guaranteed because of
two reasons: 1) In four years other Consuls will emerge; 2)
Any decision still has to be voted upon by the Governors.
This way we can concentrate on development and deliver everything
that we laid out in this paper, but at the same time, the protocol
guarantees that the system itself will definitely become more and
more decentralized and the weight of Humanode's initial group will be
diluted. Another authoritarian point is that in the first four years
of Humanode's existence proposals that require grants from
Formation must be voted upon by 66\% of the Consuls to be approved.
This precaution is taken to defend the Formation vault from many
angles of attacks that persist in decentralized permissionless public networks.

\subsection{The iron law of oligarchy}

\emph{``Who says organization, says oligarchy.''}

\emph{``Historical evolution mocks all the prophylactic measures that have
  been adopted for the prevention of oligarchy.''}

-     Robert Michels

This hypothesis was developed by the German sociologist Robert Michels
in his 1911 book, 'Political Parties.' It states that any organizational
form inevitably leads to oligarchy as an 'iron law'.
Michels researched the fact that large and complex organizations
cannot function efficiently if they are governed through direct democracy.
Because of this, power within such organizations is always
delegated to a group of individuals.

In Michels's understanding, any organization eventually is run by a
class of leaders regardless of their morals or political stance.
Monarchies and republics, democracies and autocracies,
political parties, labor unions, and corporations, etc.
have a nobility class, administrators, executives, spokespersons,
or political strategists. Michels stated that only rarely do
representatives of these classes truly act as servants of the people.
In most cases, people become pawns in never-ending games of power balancing,
networking, and survival. Regardless of the inception principles,
the ruling class will always emerge and in time it will inevitably
grow to dominate the organization's power structures.
The consolidation of power occurs for many different reasons,
but one of the most common ways is through controlling access to information.

Michels argues that any decentralized attempts to verify the
credibility of leadership are predetermined to fail, as power
gives different tools to control and corrupt any process of verification.
Many different mechanisms allow serious influence on the outcome of
democratically made decisions like the media. Michels stated that
the official goal of representative democracy of eliminating elite
rule was impossible, that representative democracy is a façade
legitimizing the rule of a particular elite, and that elite rule,
which he refers to as oligarchy, is inevitable \cite{Hyland1995}.

This law is directly applied to modern elites.
The financial network is always a complex multi-layer construct
that requires a great deal of administrative and organizational power.
According to Michels, such a system would inevitably become oligarchic.
While designing the basic principles of the Humanode network and Vortex,
the Humanode core was faced with a challenge to find a delicate balance
between organizational efficiency and the democratic involvement of the masses.
We believe that a combination of voting power equality, unbiased
intellectual barriers, direct delegation, Proof-of-Time,
Proof-of-Devotion, and proof-of-human existence would make a very
balanced and just system, but it will not solve the problem of ‘Iron Oligarchy,'
as a leadership class will definitely emerge.

Fiat credit-cycle systems have large financial entities,
PoW networks are faced with miner cartels,
PoS systems have validator oligopolies, and Humanode
has Consuls and research groups. Governors have different proposal
rights based on different tiers. Consuls have absolute freedom
in proposal creation as they can put forth an idea of any type
and they wield a right to veto any decision that is approved by
Vortex twice. Legate and Consul freedom of authority is balanced
out by the voting mechanism that requires a quorum and an absolute
majority of those voting for a proposal to be approved. As the
absolute majority of Governors is required for a decision to be
approved, it negates the ability of Legates and Consuls to approve
something against the will of the majority of voters.

In a perfect world where all participants of the network actively govern,
this balancing effort should be just enough to minimize the
influence of any type of oligopoly that might emerge in the
Humanode network, but we do not live in a perfect world.
The apathy of voters is a scourge to most of the voting systems
that exist and creates the necessity of vote delegation, which
has its own advantages and disadvantages.

\subsection{Vote delegation}

Problems of vote delegation have always accompanied any large
democratic system. The core problem of democracies in their
purest form is that they are very vulnerable to the Byzantine
Generals Problem (BGP). Any system has a critical point of failure.
Large systems tend to have several or dozens. Because of this, any
democratic system requires institutions built on top to protect
those critical points. These institutions limit the direct voting
of the masses on crucial matters. There are four main reasons why
these limitations are a necessity.

\begin{enumerate}
  \item \emph{Strategic resources, critical points, and stability.}
        Any system has a sensitive part. For example, some countries wield
        nuclear arsenals and have democratic political systems.
        The vote on the deployment of nuclear weaponry is commonly
        restricted to a very small group of individuals. It makes
        sense that such an important spectrum would be heavily guarded
        against any angle of attack, especially the BGP. That is why
        this part of the system requires consolidation of power and an
        autocratic approach in decision making. Besides weapons of mass
        destruction, there are financial, energetic, military, trading,
        diplomatic, intelligence, etc. chokepoints that unless safeguarded
        can be used by the enemies of that system to cause catastrophic
        events and lead to destabilization. Natural autocracy rises in the
        chokepoints of strategic value.

  \item \emph{Apathy of voters and effectiveness.}
        Lack of caring among voters in voting procedures can lead to a
        halt in governance, as most voting requires some kind of a quorum.
        If apathy is strong enough to stop a quorum from being raised then
        the governance process stops until a quorum is reached.
        Some operations and decisions require the constant active
        involvement of voters, which is where delegation comes in hand.
        Ordinary people do not want or have time to participate in governance,
        which is why in representative democracies citizens can cast their
        vote to elect representatives that are actively involved in decision
        making. The fewer people participate in voting, the easier it is to coordinate.

  \item \emph{Technological limitations.} Before the digital era,
        there was no effective way to conduct voting procedures,
        as communications were not as developed as they are now.
        Without proper confirmation of identity and support of modern tech,
        it was hard to imagine a way to conduct large direct voting without
        putting strain on administrative resources. Delegating to a politically
        active person negates the necessity of using sophisticated technologies
        to conduct legislative procedures.

  \item \emph{Misrepresentation.} In most democracies your vote is
        restricted by the region you are geographically located in,
        meaning that you can cast a vote for a nominee tied to your constituency,
        but he might not get elected, meaning that your vote was practically
        burned and a person that you did not vote for might be representing you.
        Most governing systems lack the freedom of vote delegation, as you
        cannot directly delegate your voice to a particular person.
\end{enumerate}

While devising the voting procedures for Vortex, the Humanode core
has kept in mind the principles mentioned above. The Governor tier
system safeguards critical points by limiting the abilities of the
electorate to create proposals but at the same time, the autocratic
chokepoint is balanced out by requiring a quorum of Governors to
approve created proposals. The influence of apathy of voters is limited by
demanding voting activity from human nodes to be counted as Governors.
This way only active participants of the network are counted in reaching a quorum.
The technological progress in DAO deployment and biometric processing in
the last decade has brought forward a way to overcome the obstacles of
the past connected to direct voting procedures and the uniqueness of voters.
Delegation of voting power is permissionless, meaning that any human node
can delegate its vote to any Governor in the Humanode network. We acknowledge
that even with modern approaches to voting and technological breakthroughs, a
delegation mechanism in the Humanode network is a natural necessity.

The digital revolution has paved the way for technologies that allow
us to create systems with liquid representative democracies. Compared to
traditional representative democracies, a voter can re-cast his vote any time
he wants, without the necessity to wait for years to do it again.
Vote delegation can be changed anytime. Delegated PoS (DPoS) protocols
implemented liquid democracy for delegating transaction validation operations
to professional entities. As the validators are safeguarding the protocol
and receive a commission for their operation, the voter's choice is usually
driven by economic incentives: how the commission size, uptime, and
security of the delegate's server might reflect on the voter's earnings.
Is that enough to choose an opinion representative in a decentralized
network? Most DPoS networks have a strict unbounding period that can
last up to two weeks or even months. This measure is a necessity to
safeguard the system from manipulated panic-based market crashes where
Delegators undelegate their tokens and sell them in fear of losing value.
In the Humanode network, voting power is not entangled with a token, which
is why there is no need for unbounding periods. Any time a human node wishes
to re-cast or simply retrace its delegation it can be done instantly.

\subsection{Populist tide and professional backslide}

It is commonly acknowledged that any voting system faces the
problem of too much populism. Hypothetically there are two major
approaches to how populism is perceived:

\begin{itemize}
  \item Populism poses a threat to democratic stability.
        According to recent studies, conducted by Jordan Kyle
        and Yascha Mounk of the Tony Blair Institute for Global Change,
        one of the key findings they have had is that populists are far
        more likely to damage democracy. Overall, 23 percent of populists
        cause significant democratic backsliding, compared with 6 percent of
        non-populist democratically elected leaders (J. Kyle \& Y. Mounk, 2018).
        In other words, populist governments are about four times more
        likely than non-populist ones to harm democratic institutions.

  \item Populism is a necessary corrective mechanism that addresses
        popular problems and limits the power of elites.
\end{itemize}

Regardless of which view is more accurate, populism is
acknowledged to be a very powerful tool to gather the support
of the masses in democratic systems. The main danger perceived
by the Humanode core is the rise of populists. Individuals that
know how to be popular do not necessarily have the intelligence,
professional qualities, experience, or profound knowledge on the
subjects they have to make decisions upon on a regular basis.

In the Humanode network, every human node has a voting power of 1.
Voting delegation in Humanode allows for any human node to delegate their
voting power to any Governor in the network. Governor power equals $1 +$
the number of delegations from other human nodes. Such a system allows
limitless crowdsourcing possibilities as delegation is liquid and not
regionally bound. As in any other democratic system, individuals that
possess oratory, diplomatic skills and are backed by influential
media sources have an advantage in the Humanode network.
An introvert with sociopathic tendencies possessing a very professional
skill set for decision-making operations will most likely receive less
support than a good negotiator, orator, and crowd controller that possesses
a mediocre skill set. This is slightly balanced out by the fact that human
nodes must have an accepted proposal before they become Governors.
Thus Governors should be less affected by populist media, as they
have a confirmed intellectual skill set that allowed them to create a
useful proposal accepted by the Governors of Humanode.

In Vortex voting procedures, Governors have disproportionate
voting power and those Governors that have more delegations have more power.
The professional backslide in our understanding poses a threat to the
effectiveness, progressiveness, and constant optimization of governance.
We fear that without Proof-of-Devotion, which is in a way a proof of
having some kind of professional skill set, any democratic system faces
becoming a plutocracy, where the wealthiest members control influential
and credible media sources to direct the opinion of masses and drive
support to candidates of their choosing.

Proof-of-Devotion might bring a small balance to populism upheaval,
as it demands participation in Formation to receive proposal rights
on critical matters. Nevertheless, Consuls wielding huge delegations
will inevitably emerge and their stance in decision-making mechanisms
will be very strong. The only way to limit their influence is the direct
and active participation of human nodes in governing processes.
The more Governors that do not delegate their vote and actively participate in
governance the less authority can be accumulated in the hands of those that seek it.

\bibliography{references}
\bibliographystyle{ieeetr}

\end{document}